\documentclass[journal]{IEEEtran}

\ifCLASSINFOpdf
\else
    \usepackage[dvips]{graphicx}
\fi

\usepackage{subfigure}
\usepackage{multirow}
\usepackage{multicol}
\usepackage[dvipdfm,
            colorlinks=true,
            bookmarks=false,
            citecolor=black,
            linkcolor=black,
            anchorcolor=black,
            urlcolor=black 
            ]{hyperref}
\usepackage{array}
\usepackage{slashbox}
\usepackage{color}

\begin{document}

\title{Identification of Image Operations Based on Steganalytic Features}

\author{
Haodong~Li,
Weiqi~Luo,
Xiaoqing~Qiu,
Jiwu~Huang
\thanks{W. Luo (corresponding author) is with the School of Software, Sun Yat-sen University, Guangzhou 510006, China. (e-mail: luoweiqi@mail.sysu.edu.cn).}
\thanks{H. Li and X. Qiu are with the School of Information Science and Technology, Sun Yat-sen University, Guangzhou 510006, China. (e-mail: lihaod, qiuxq3@mail2.sysu.edu.cn).}
\thanks{J. Huang is with the College of Information Engineering, Shenzhen University, Shenzhen 518060, China. (e-mail: jwhuang@szu.edu.cn).}
\thanks{Part of this work was presented at the 19th IEEE International Conference on Image Processing (ICIP '12) and the 2nd ACM Workshop on Information Hiding and Multimedia Security (IH\&MMSec '14).}
}

\markboth{}%
{}

\maketitle

\begin{abstract}
Image forensics have attracted wide attention during the past decade. Though many forensic methods have been proposed to identify image forgeries, most of them are targeted ones, since their proposed features are highly dependent on the image operation under investigation. The performance of the well-designed features for detecting the targeted operation usually degrades significantly for other operations. On the other hand, a wise attacker can perform anti-forensics to fool the existing forensic methods, making countering anti-forensics become an urgent need. In this paper, we try to find a universal feature to detect various image processing and anti-forensic operations. Based on our extensive experiments and analysis, we find that any image processing/anti-forensic operations would inevitably modify many image pixels. This would change some inherent statistics within original images, which is similar to the case of steganography. Therefore, we model image processing/anti-forensic operations as steganography problems, and propose a detection strategy by applying steganalytic features. With some advanced steganalytic features, we are able to detect various image operations and further identify their types. In our experiments, we have tested several steganalytic features on 11 different kinds of typical image processing operations and 4 kinds of anti-forensic operations. The experimental results have shown that the proposed strategy significantly outperforms the existing forensic methods in both effectiveness and universality.
\end{abstract}


\section{Introduction}
\label{Sec:Introduction}

With the rapid development of image processing techniques, digital images can be easily modified without leaving any perceptible artifacts. Digital image forgeries are now abused in our daily life, leading to potential serious moral, ethical, and legal consequences. Therefore, image forensics \cite{Stamm2013} have attracted increasing attention. Up to now, many forensic methods have been proposed. The existing forensic methods assume that there are some inherent statistics within original natural images. Such statistics vary according to different image sources, and they would change after different operations. By gathering features to characterize the image generation pipeline, one can determine the source camera \cite{Lukavs2006a} or identifying the camera model \cite{Swaminathan2007} for a given image. By analyzing the special artifacts left by a certain image operation, it is possible to find out the tampered images. For instance, identifying the JPEG compression history \cite{Fan2003,Farid2009,Luo2010,Bianchi2012}, revealing contrast enhancement \cite{Stamm2008,Cao2014}, resampling \cite{Popescu2005,Mahdian2008,Li2013} and median filtering \cite{Kirchner2010,Yuan2011,Kang2012,Chen2013a}, exposing image splicing \cite{Shi2008,He2012,Zhao2015}, and so on. The employed features in most of existing methods, however, are usually specially designed for only one type of operation, and thus they are difficult to be generalized for detecting other operations. Besides, many methods assume that the suspected image has been processed either by a specific operation or not, which means that just a binary classification is considered. However, such an assumption seems not very reasonable in practice, since the previous operations for a given suspected image are usually known. In such a case, we can not decide which pre-designed binary classifier should be applied.


Another serious issue that would affect the performances of the existing forensic methods is the presence of anti-forensics\cite{Boehme2013}. In order to fool the existing forensic methods, a wise attacker can perform some anti-forensic operations to weaken the artifacts left by tampering or confuse the inherent statistics within original images. So far, there have been several attempts to carry out anti-forensics against forensic detection, such as \cite{Stamm2011a,Fan2014,Cao2010,Kwok2011,Kirchner2008,Wu2013}.
At this point, it is urgent to expose the anti-forensically altered images. In \cite{Valenzise2011,Valenzise2013}, Valenzise \textit{et al.} proposed a method to detect anti-forensics of JPEG compression \cite{Stamm2011a} by measuring image quality with the total variation. In \cite{Lai2011}, the authors presented two detectors to reveal the JPEG anti-forensic operation.
Cao \emph{et al.} \cite{Cao2012} introduced a semi-nonintrusive approach to detect anti-forensics of resampling \cite{Kirchner2008} through analyzing the output of resampling software with some specifically designed images. Zeng \emph{et al.} \cite{Zeng2014} identified anti-forensics of median filtering \cite{Wu2013} via detecting the peaks in Fourier transform domain for the ratio of horizontal pixel differences with zero values. Similar to the targeted forensic methods that aim at detecting a certain operation, though the counter anti-forensic methods can effectively expose the resulting images after some anti-forensic operations, their applicability is still limited since each of them just focuses on a specific anti-forensic operation.

In our previous work \cite{Li2012}, we employed JPEG steganalytic features \cite{Chen2008a} to detect JPEG anti-forensic operation \cite{Stamm2011a}. Though only anti-forensics of JPEG compression was considered, it shows the possibility of detecting anti-forensics with some steganalytic features. In the subsequent work \cite{Qiu2014}, we use steganalytic features to detect image splicing and some common image operations, and further identify the types of them.
As the extension of our previous works, this paper considers the task of identifying both various image processing and anti-forensic operations, and tries to find a universal feature to deal with such problem. Via analyzing various image operations (in this paper, image operations limit common image processing and most existing anti-forensic operations), we find that any operation would modify many pixel values and thus inevitably destroy some inherent statistics of original images, which is similar to the process of data hiding. Via modeling image forensics as a steganalytic problem, we present a strategy to identify various image operations with some steganalytic features. Compared with \cite{Li2012} and \cite{Qiu2014}, the main differences and extensions in this paper are as follows. First, instead of just presenting some preliminary experimental results in our previous works, we give a detailed comparison between image operations and steganography to make the motivation and the presented idea more clear, and further study the feasibility on using steganalytic features to detect various image operations (refer to Section \ref{sec:Detection}). Second, more image operations are analyzed and evaluated in our experiments, including 11 kinds of typical image processing operations and 4 kinds of existing anti-forensic methods. The extensive experimental results have shown that the proposed strategy works much better than the existing forensic methods in both effectiveness and universality (refer to Section \ref{Sec:Results}).

Compared with previous arts about forensics and/or detecting anti-forensics, this paper has some differences and new insights for forensics as follows:
\begin{itemize}
\item Unlike those targeted  methods that aim at detecting a specific operation (e.g., \cite{Luo2010,Cao2014,Mahdian2008,Valenzise2011}), we need not to carefully analyze the specific traces left by the operation under investigation. Instead, we use exactly the same features for detecting various different operations, which means that the steganalytic features employed in this paper can be regarded as universal ones.

\item Though a few literatures such as \cite{Shi2008,Boehme2013} have discussed the similarities between steganalysis and forensics, they just provide some qualitative analysis. In this paper, we further demonstrate the correlations between steganalysis and forensics with detailed examples and extensive quantitative analysis (refer to Section \ref{subsec:vsSteg}), and deeply assess the suitability of that applying steganalytic features for forensic detection (refer to Section \ref{subsec:visualization}).

\item It is noted that applying steganalytic features in forensics is not new, a few prior works have used some steganalytic features or extracted features analogous to steganalysis. For instance, SPAM (subtractive pixel adjacency matrix) \cite{Pevny2010} was used in \cite{Kirchner2010} for median filtering detection, features based on rotation invariant local binary pattern \cite{Ojala2002} were respectively used in \cite{Xu2012,Ding2014} for camera model detection and image sharpening detection; SRM (spatial rich model) \cite{Fridrich2012} was used in \cite{Cozzolino2014} for detecting image splicing. However, all of them just focus on detecting one specific operation. Unlike the previous works, we systematically study whether the advanced steganalytic features can be universally used for detecting various image operations, and demonstrate their universality and effectiveness with extensive experiments. Furthermore, we first try to identify the type of operation via multiple classification (refer to Section \ref{subsec:classification} and \ref{subsubsec:multiclass}), which is never considered before yet very important in practice.
\end{itemize}

The rest of this paper is organized as follows. Section \ref{sec:Detection} analyzes the commons of various image operations from the view of steganalysis, and proposes a strategy to detect different operations via applying steganalytic features. Section \ref{Sec:Proposed} presents the implementation of the proposed detection strategy, including feature selection and design of classifiers. Section \ref{Sec:Results} shows the experimental results and discussions. Finally, the concluding remarks are given in Section \ref{sec:Conclusion}.

\begin{figure}[t]
  \centering
  \subfigure[Spatial correlations]{\includegraphics[scale=0.85]{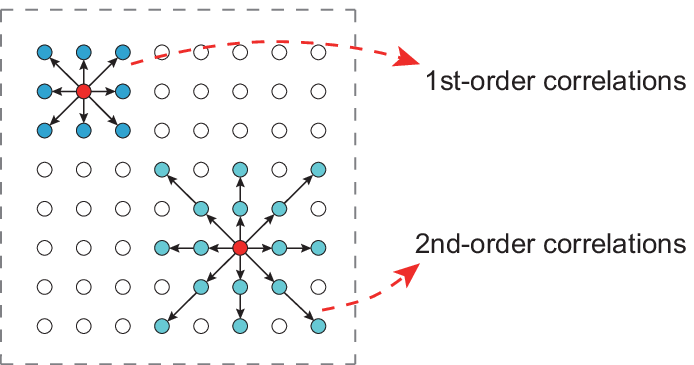}}
  \subfigure[Frequency correlations]{\includegraphics[scale=0.85]{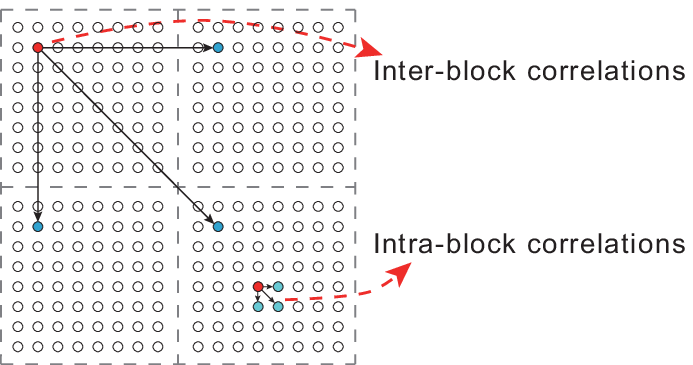}}
  \caption{Illustrations of the correlations of local image regions in spatial domain and DCT frequency domain.}
  \label{fig:img_corr}
\end{figure}

\section{Analysis on image processing and anti-forensic operations}
\label{sec:Detection}

In this section, we firstly analyze the commons of various image processing operations and existing anti-forensic operations, and then introduce a strategy for detecting them from the view of steganalysis.

\subsection{Pixel Modification in Various Image Operations }
\label{subsec:modification}

\begin{table*}[!t]
\centering
\caption{The average modification ratios and PSNR after performing various operations on 10,000 images in Boss Base image database with different parameters (please refer to Table \ref{table:parameter} for more details). ``AF'' and ``CE'' denote ``anti-forensics'' and ``contrast enhancement'', respectively. }{
 \begin{tabular}{|c|c|c||c|c|} \hline
    \multicolumn{3}{|c||}{Operation} & Modification & PSNR \\ \hline \hline
                              & \multicolumn{2}{c||}{Contrast enhancement} & 99.17\% & 16.06 dB \\ \cline{2-5}
    Image processing                    & \multicolumn{2}{c||}{Sharpening}          & 73.91\% & 36.77 dB \\ \cline{2-5}
    operations                 & \multicolumn{2}{c||}{Spatial Filtering}           & 79.11\% & 34.45 dB \\ \cline{2-5}
                              & \multicolumn{2}{c||}{Lossy compression}   & 68.33\% & 43.90 dB \\ \hline
    & \multirow{2}{*}{JPEG AF} & Dither \cite{Stamm2011a}\protect\footnotemark                 & 81.56\% & 39.49 dB \\ \cline{3-5}
    \multirow{2}{*}{Anti-forensic}&  & Dither \& deblocking \cite{Stamm2011a}   & 85.87\% & 35.10 dB \\ \cline{2-5}
    \multirow{2}{*}{operations}& \multirow{2}{*}{CE AF}
                               & Cao's method \cite{Cao2010}              & 99.37\% & 20.20 dB \\ \cline{3-5}
    &                          & Kwok's method \cite{Kwok2011}            & 99.37\% & 20.21 dB \\ \cline{2-5}
    & \multicolumn{2}{c||}{Median filtering AF \cite{Wu2013}}             & 86.35\% & 34.11 dB \\ \hline
\end{tabular}}
\label{tab:modifications}
\end{table*}

\begin{figure*}[t]
  \centering
  \includegraphics[scale=.95]{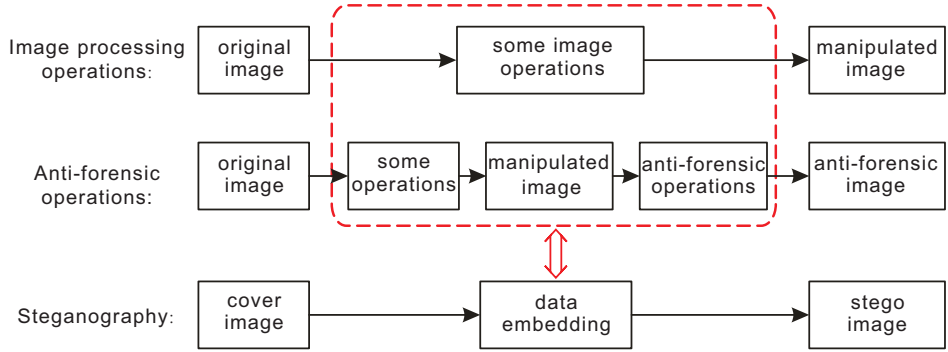}
  \caption{Image processing operations and anti-forensic operations vs. steganography.}
  \label{fig:AntivsEmb}
\end{figure*}

To seek a universal feature set for detecting various image operations, we mean to obtain a feature set that can capture the artifacts left by various operations rather than a specific one that only works for a certain operation. Based on our extensive experiments and analysis, we found that lots of pixel values would be modified after various operations, as shown in Table  \ref{tab:modifications} (here we do not consider the operations that are relevant to resampling due to lack of synchronization). It is observed that more than 70\% pixel values are modified. Besides, the quality of the resulting images also degrades significantly due to the modifications. All the average PSNR (Peak Signal-to-Noise Ratio) of the resulting images relative to their original versions are lower than 44 dB. Especially, the average PSNR for the contrast enhancement operations including Gamma correction and histogram equalization drops to about 16 dB.

As we know, the adjacent pixel values and frequency coefficients within an original natural image are highly dependent, as illustrated in Fig. \ref{fig:img_corr} (a) and (b). It is expected that once an image pixel is modified, it would inevitably affect the relationship with its neighbors. Thus such inherent correlations are very difficult to be well preserved after pixel modifications especially when the quantity of modified pixels are large. Usually, the more pixels are changed, the easier the resulting images can be detected. Furthermore, various image operations modify the pixels within an original image in different manners and/or strengths, which means that they would destroy the inherent correlations among adjacent pixels in different degrees. By properly measuring such correlations, it is possible to distinguish the manipulated images from the original natural ones, and  further identify the type of various operations.

\subsection{Various Image Operations vs. Steganography}
\label{subsec:vsSteg}

Based on previous observations in Subsection \ref{subsec:modification}, various image operations would modify many image pixels and destroy the inherent correlations among adjacent pixels in different degrees. Thus how to model the correlations in natural images is the key issue for detecting various image operations. Fortunately, many useful statistical models in the field of steganalysis can be adopted due to some similarities between steganography and various image operations.

\footnotetext{The anti-forensic method proposed in \cite{Stamm2011a} consists of two steps, \emph{i.e.}, adding dithers to conceal the quantization artifacts, and subsequently applying deblocking to remove the blocking artifacts. These two operations are denoted as ``dither'' and ``dither \& deblocking'', respectively.}

As illustrated in Fig. \ref{fig:AntivsEmb}, both image processing and anti-forensic operations are similar to the process of data embedding in steganography, since they all have to modify some pixel values in original (cover) images. As described previously, pixel modifications would destroy the inherent correlations among adjacent pixels within the image.
In order to detect the pixel modifications in steganography, many steganalytic features have been proposed via modeling such inherent properties. These features are usually effective even for relatively low modification rates. For example, the literature \cite{Fridrich2012} has shown the effectiveness when the modification rate is around 9\% (0.40 bpp for WOW \cite{Holub2012}).
Therefore, we are wondering whether the steganalytic features can be also effective for identifying image processing and anti-forensic operations, which modify original images in a much more severe way. In the following, we will provide some quantitative analysis to show the  feasibility on applying steganalytic features in image forensics.

\begin{figure*}[t]
  \centering
  \includegraphics[scale=0.6]{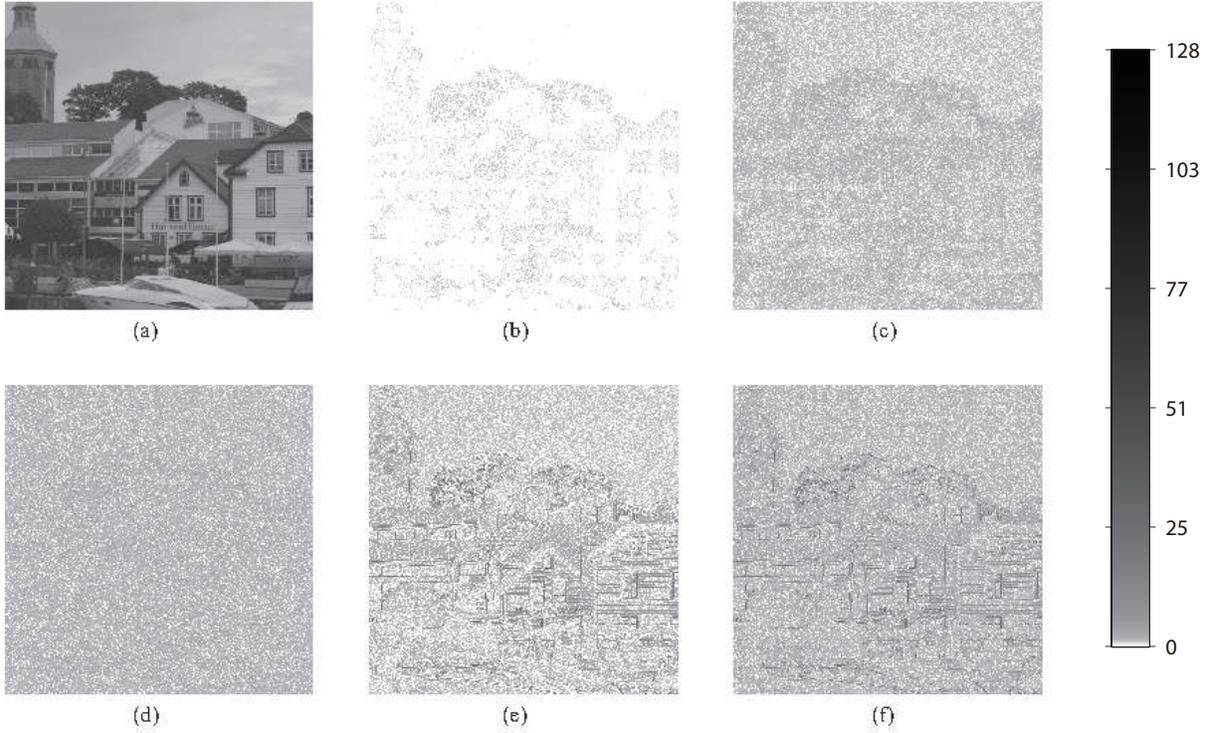}
  \caption{Illustration of the modification areas. (a) The original image. (b) Absolute differences of the stego image with WOW \cite{Holub2012} at a high embedding rate of 0.4bpp. (c) Absolute differences of the modified image with JPEG compression (QF=90). (d) Absolute differences of the modified image with anti-forensics of JPEG compression (Method \cite{Stamm2011a}, QF=90). (e) Absolute differences of the modified image with median filtering (filter size 3$\times$3). (f) Absolute differences of the modified image with anti-forensics of median filtering (Method \cite{Wu2013}, filter size 3$\times$3, $B=4,T=4$). The scale bar on the right shows the degrees of (b)-(f). The ratios of non-zero values (\textit{i.e.} the modification rates) for (b)-(f) are 8.45\%, 77.07\%, 93.60\%, 68.39\%, 88.70\%, respectively. The maximum value of absolute differences for (b)-(f) are 1, 17, 21, 108, 109, respectively.}
  \label{fig:difference}
\end{figure*}

\begin{table*}[t]
\centering
\caption{Comparison between steganography and anti-forensics evaluated on 10,000 images in Boss Base image database}{
 \begin{tabular}{|c||c|c|} \hline
     & Modern adaptive steganography & Image processing and anti-forensics operations\\ \hline \hline
     Modified & Mainly located at the  & Both textural and smooth \\
     regions  & textural/noisy regions & regions are changed \\ \hline
     \multirow{2}{*}{Modification} & \multirow{3}{*}{$\pm$1} & The absolute magnitudes are \\
     \multirow{2}{*}{magnitude}    &                         & usually much greater than 1 \\
                                   &                         & in all existing techniques \\\hline
     \multirow{2}{*}{Modification} & WOW\cite{Holub2012} - around 8.93\% & \multirow{2}{*}{Larger than 68\% on average} \\
     \multirow{2}{*}{rate}& HUGO\cite{Pevny2010a} - around 10.26\% & \multirow{2}{*}{(refer to Table \ref{tab:modifications})}\\
     & (high embedding rate: 0.4bpp) & \\ \hline
\end{tabular}}
\label{tab:compare}
\end{table*}

It is well known that there are several important factors that would significantly affect the detectability of hidden data in steganography. The first one is the location of the modified pixels. The recent literatures \cite{Holub2012,luo2010edge,Pevny2010a} have shown that compared to smooth regions in an image, textural regions are more suitable for hiding data since they are difficult to be modeled, and our human eyes are insensitive to the modifications in these regions. The second factor is the quantity and intensity of modified pixels. Based on previous studies, we know that the larger the quantity and/or intensity, the easier the stego images can be detected. To compare image operations and steganography in both factors, we illustrate the modifications of the resulting images after performing some image operations and steganography in Fig. \ref{fig:difference}. It is observed that image operations would modify both the smooth regions and the textural regions, and much more pixels are changed compared to steganography. What is more, the modification intensity for image operations is also much greater, please refer to the gray levels of the difference images as shown in Fig. \ref{fig:difference} (b)-(f).

To further provide more convincing evidences about the image operations and steganography, we evaluate 10,000 images from Boss Base v1.01 \cite{Bas2011} and summarize the results on the above two factors in Table \ref{tab:compare}. It is observed in Table \ref{tab:compare}that image operations would modify images in a much more serious manner compared to the steganography. Based on the above experiments and analysis, therefore, it is expected that some advanced statistical features used in steganalysis are suitable and easy for detecting various image operations.

\subsection{Visualizing the Pixel Differences}
\label{subsec:visualization}

In this subsection, we will use a simple steganalytic feature (i.e. the joint probability of the backward and forward differences of the image pixels) to describe the correlations among adjacent pixels, and illustrate that various image operations would change such a simple feature in different manners and/or degrees.

\begin{figure*}[t]
  \centering
  \subfigure[]{\includegraphics[scale=0.375]{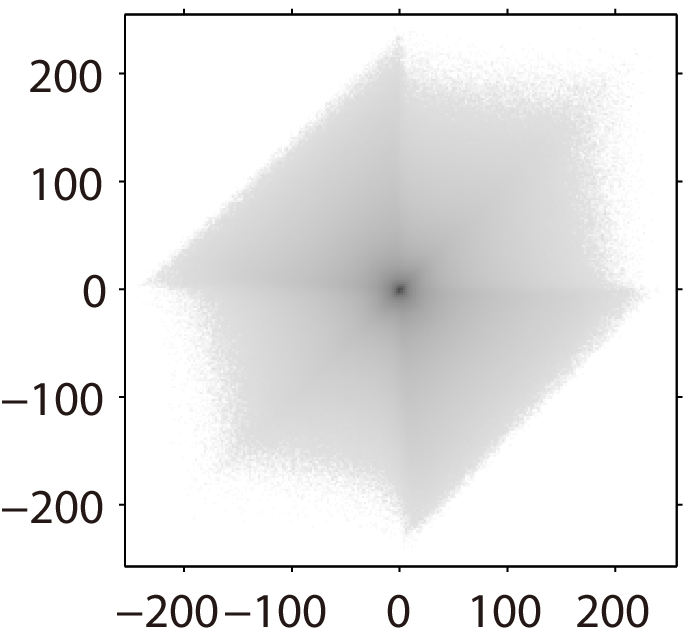}}\hspace{1em}
  \subfigure[]{\includegraphics[scale=0.375]{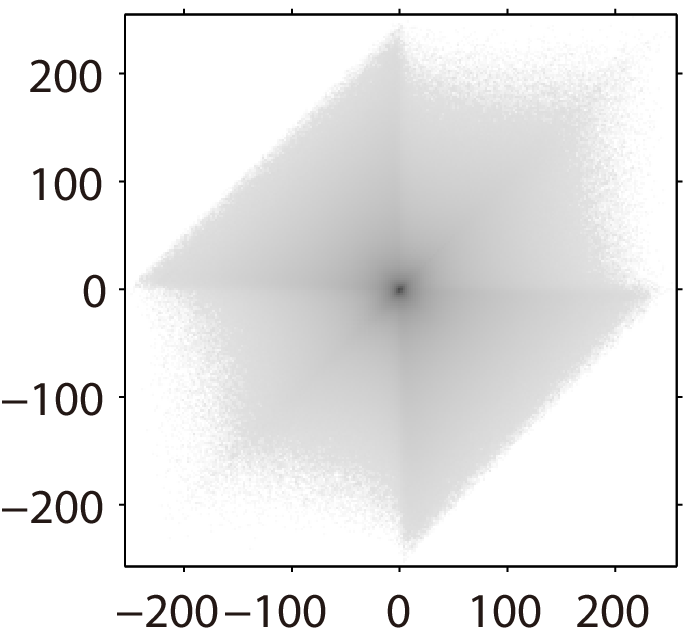}}\hspace{1em}
  \subfigure[]{\includegraphics[scale=0.375]{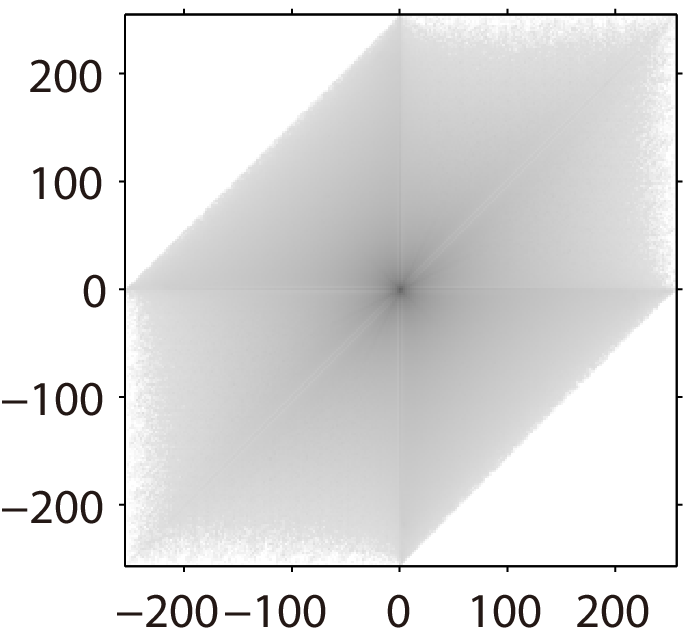}}\\
  \subfigure[]{\includegraphics[scale=0.375]{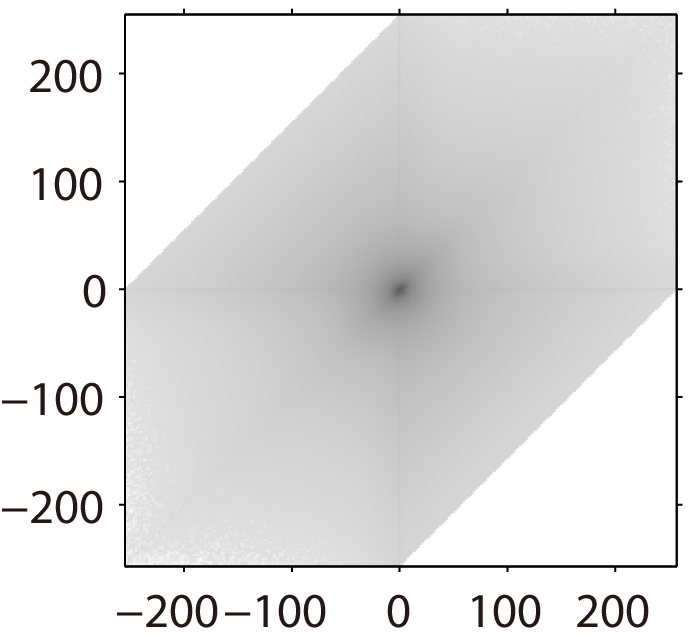}}\hspace{1em}
  \subfigure[]{\includegraphics[scale=0.375]{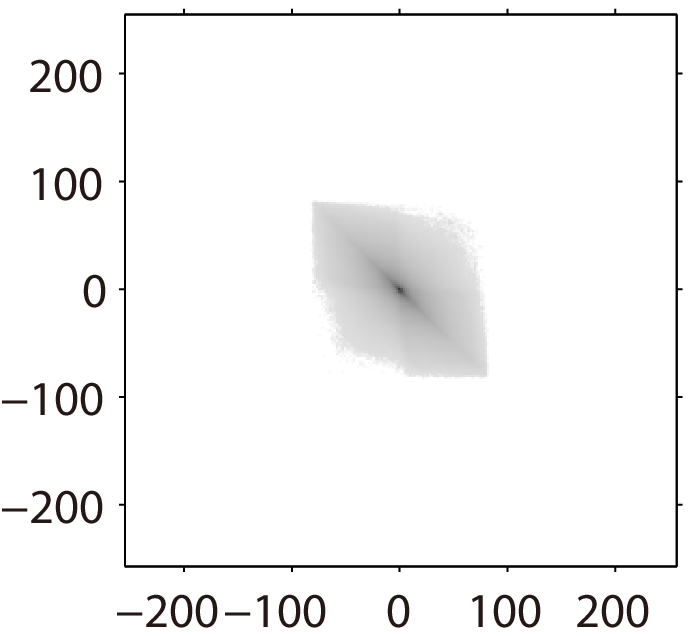}}\hspace{1em}
  \subfigure[]{\includegraphics[scale=0.375]{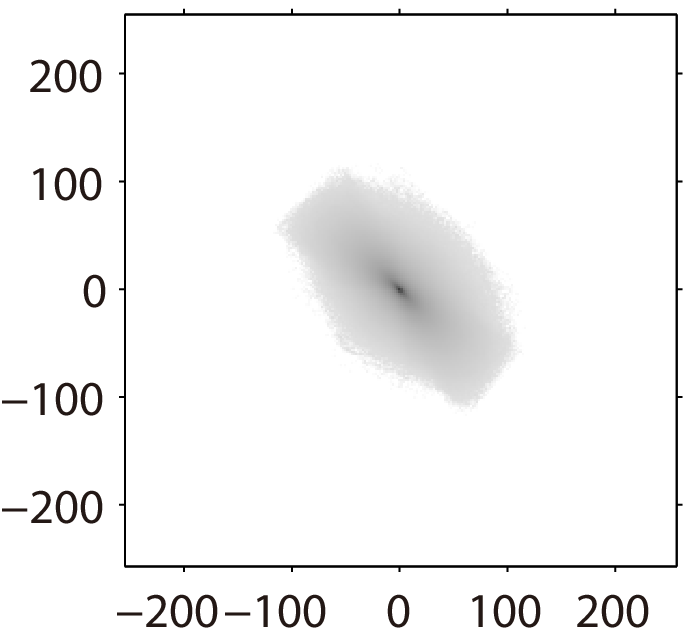}}\hspace{1em}
  \subfigure[]{\includegraphics[scale=0.375]{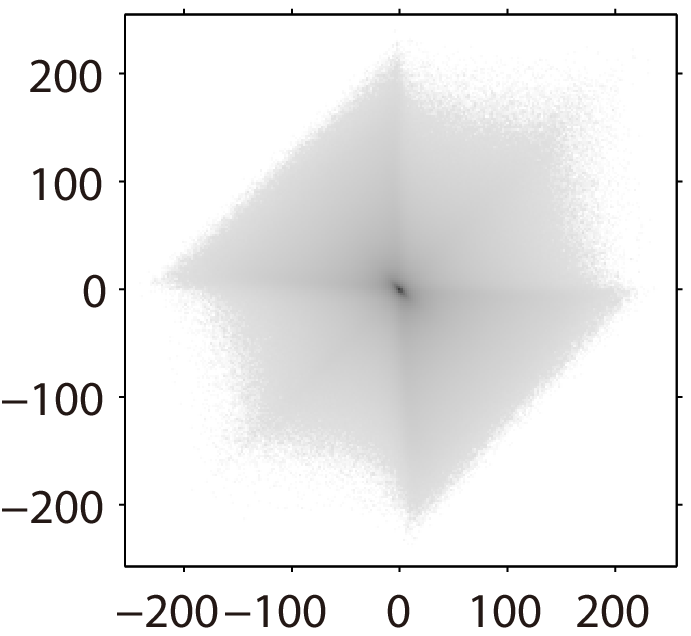}}\hspace{1em}
  \subfigure[]{\includegraphics[scale=0.375]{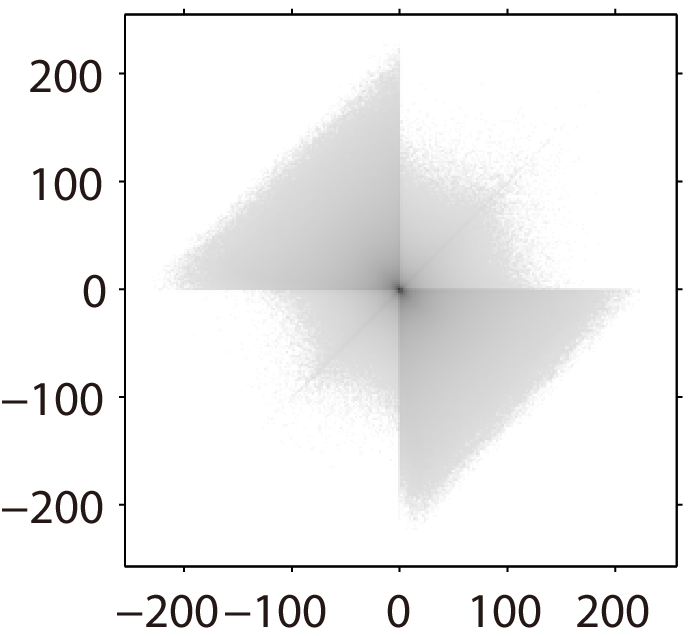}}\\
  \subfigure[]{\includegraphics[scale=0.375]{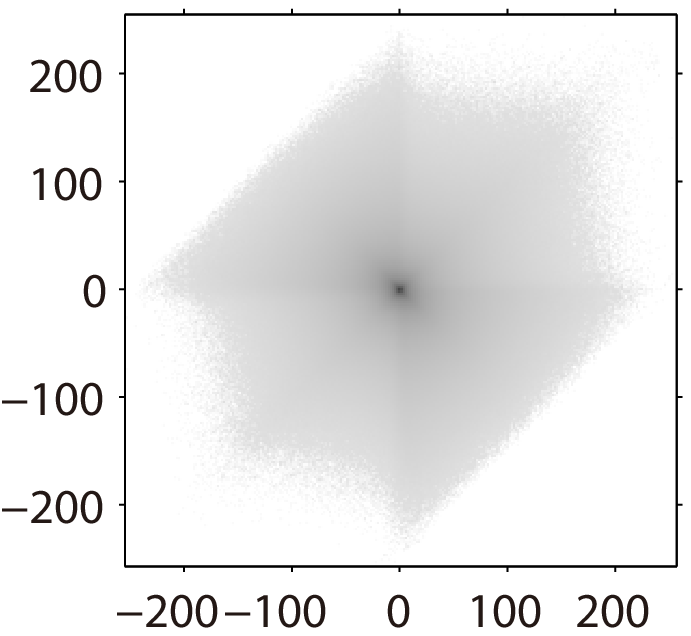}}\hspace{1em}
  \subfigure[]{\includegraphics[scale=0.375]{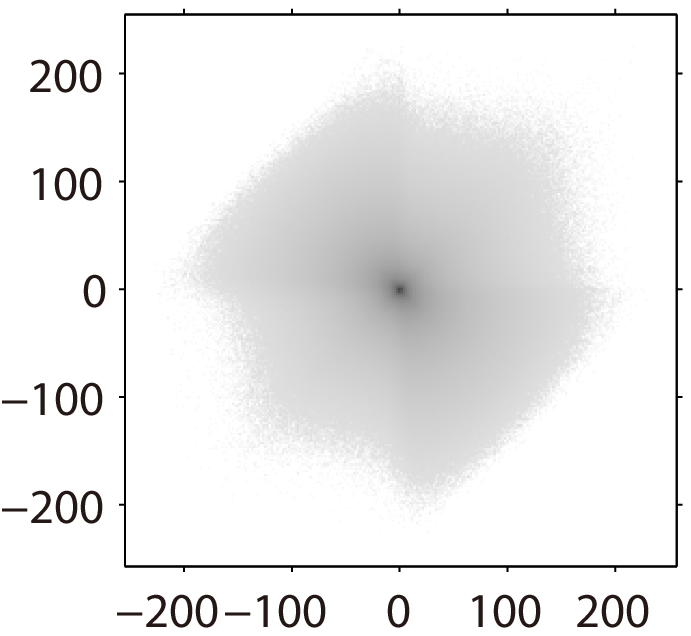}}\hspace{1em}
  \subfigure[]{\includegraphics[scale=0.375]{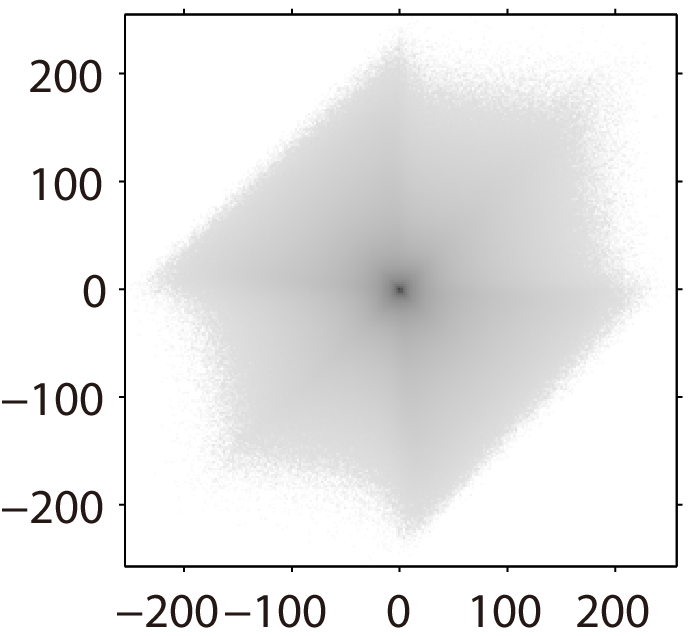}}\hspace{1em}
  \subfigure[]{\includegraphics[scale=0.375]{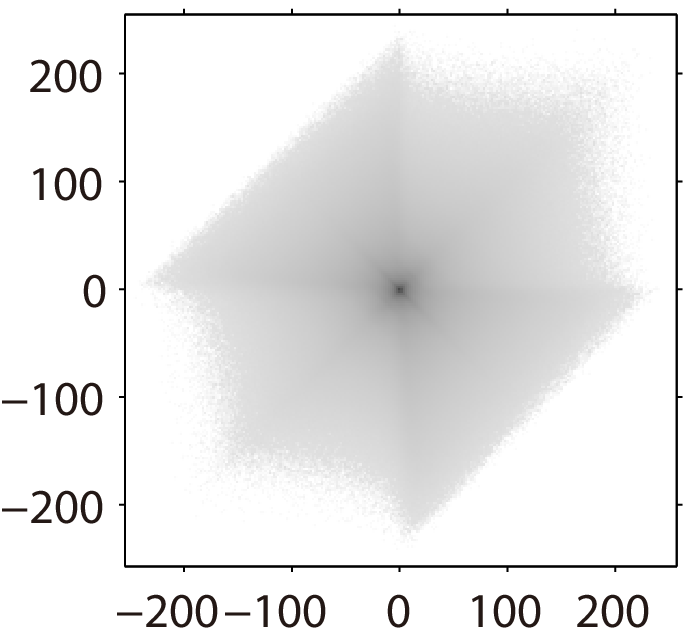}}\hspace{1em}
  \subfigure[]{\includegraphics[scale=0.375]{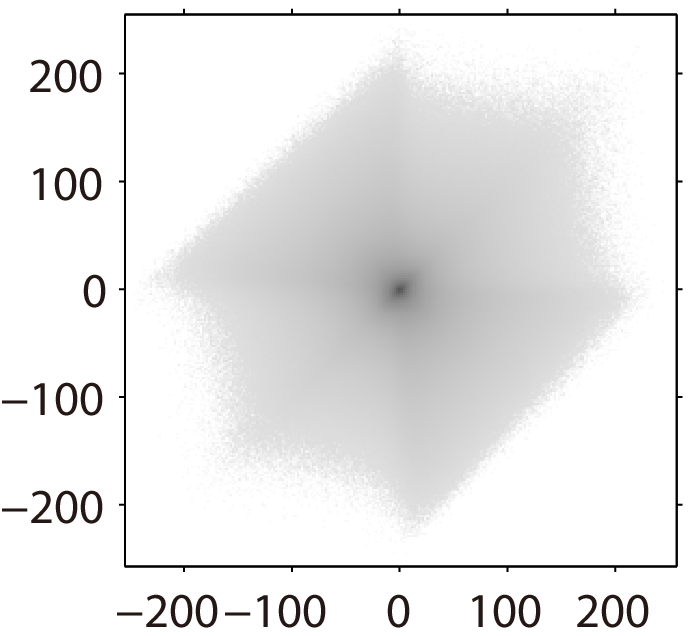}}\\
  \subfigure[]{\includegraphics[scale=0.375]{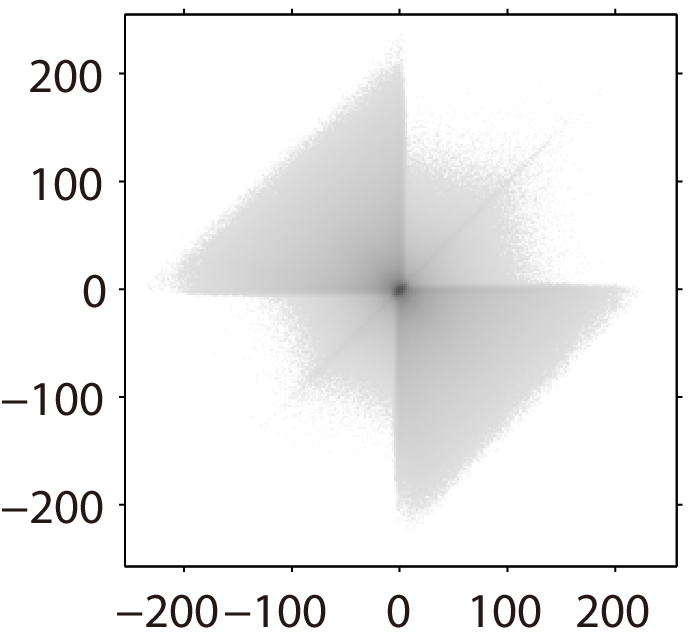}}\hspace{1em}
  \subfigure[]{\includegraphics[scale=0.375]{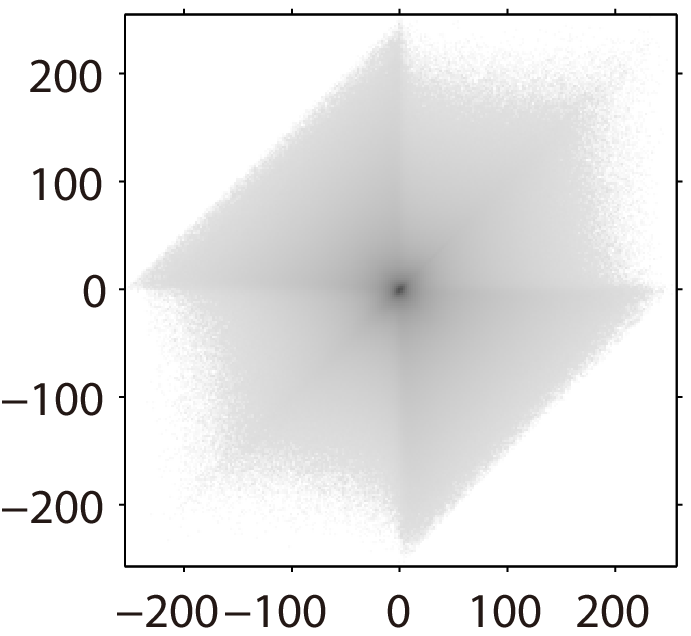}}\hspace{1em}
  \subfigure[]{\includegraphics[scale=0.375]{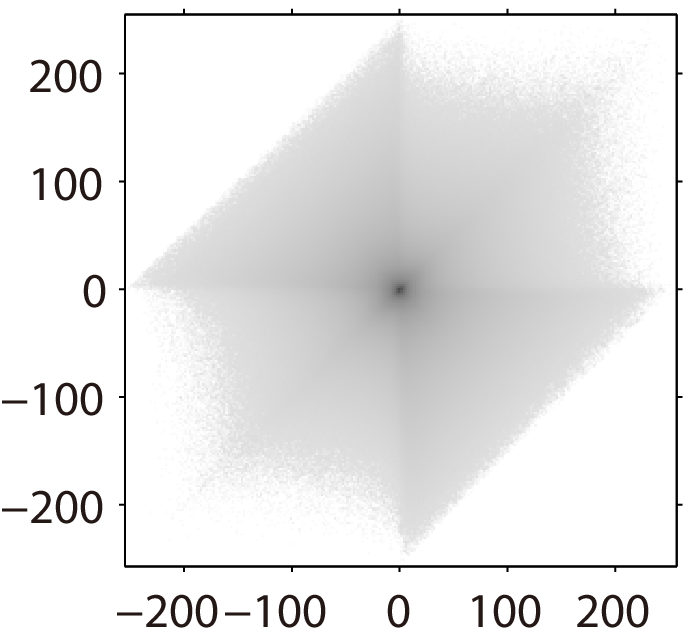}}\hspace{1em}
    \subfigure[]{\includegraphics[scale=0.375]{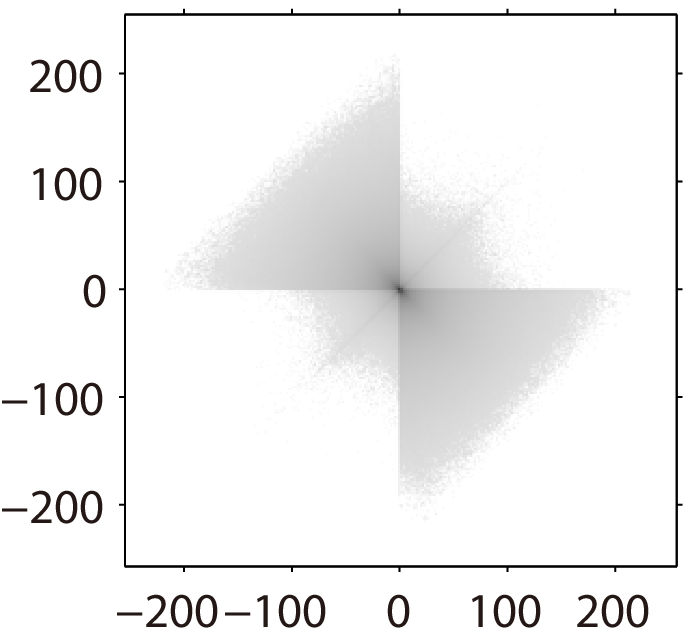}}\hspace{1em}
  \subfigure[]{\includegraphics[scale=0.375]{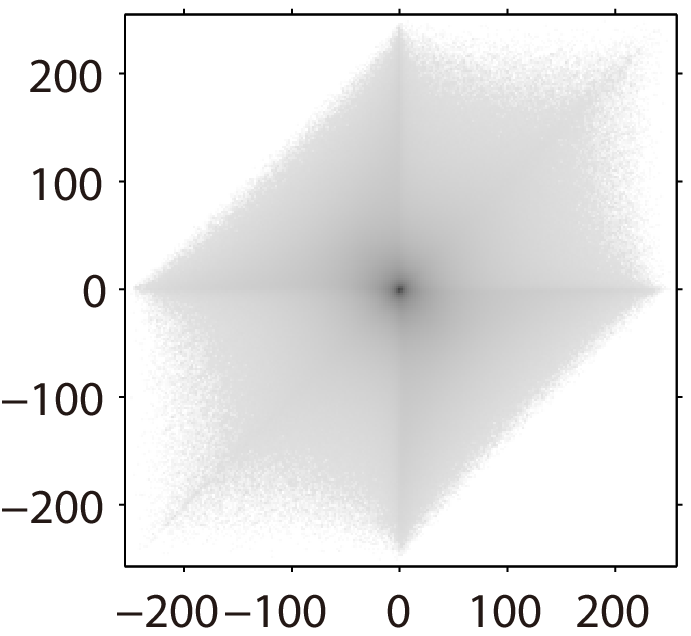}}\\
  \subfigure[]{\includegraphics[scale=0.375]{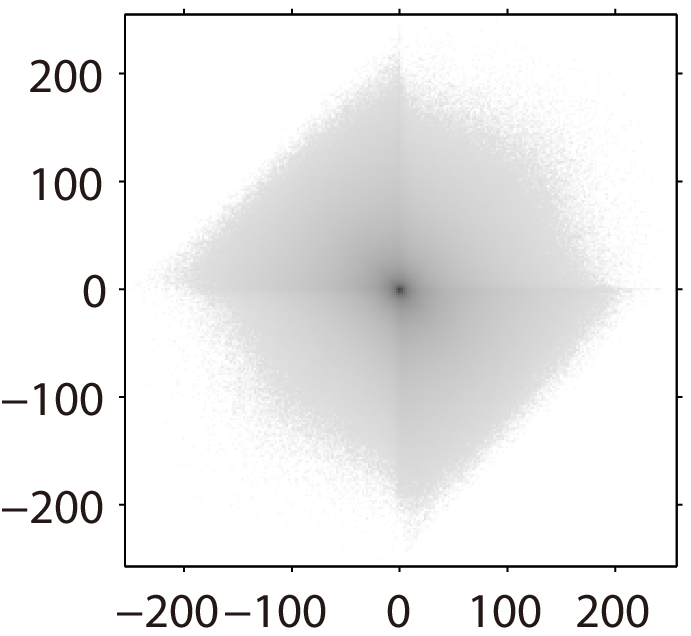}}\hspace{1em}
  \subfigure[]{\includegraphics[scale=0.375]{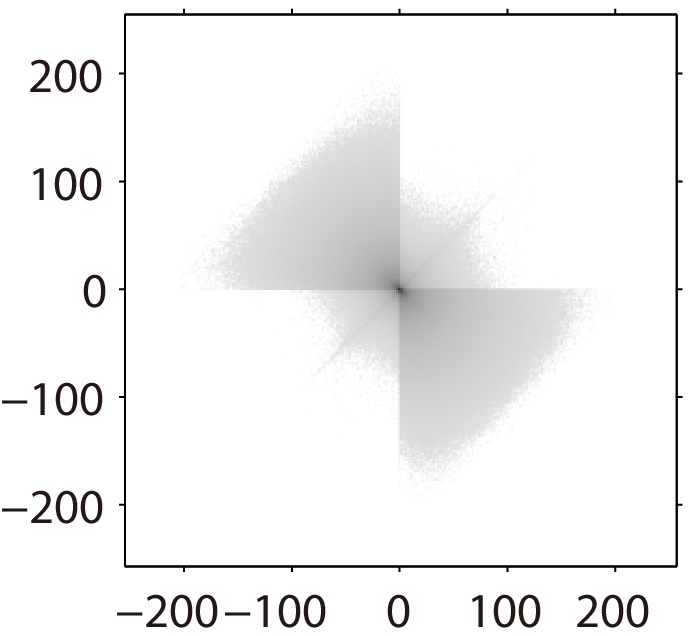}}\hspace{1em}
  \subfigure[]{\includegraphics[scale=0.375]{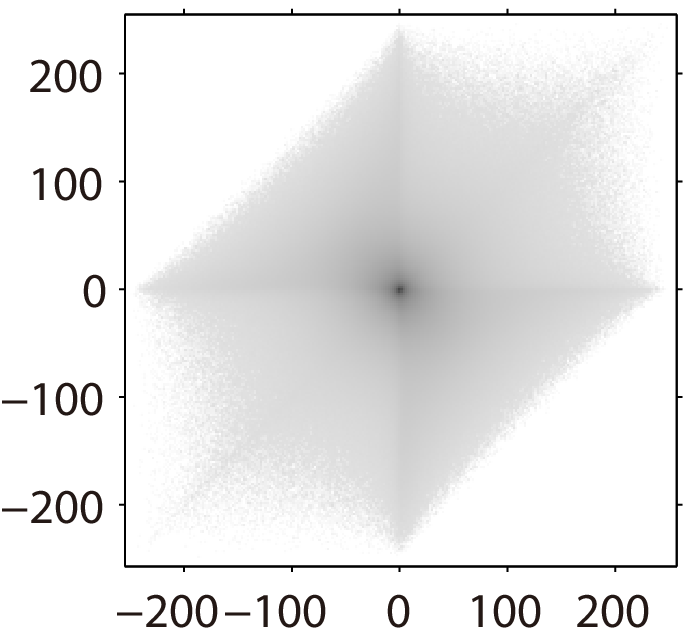}}\hspace{1em}
  \subfigure[]{\includegraphics[scale=0.375]{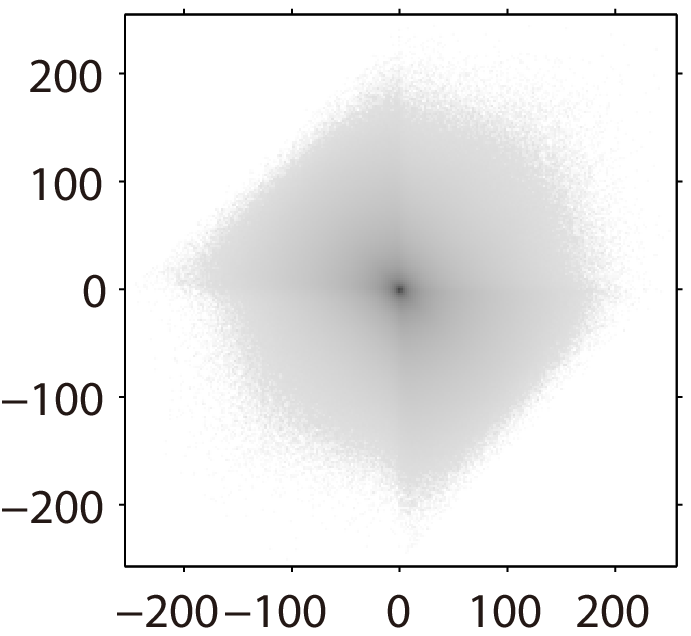}}\hspace{1em}
  \subfigure[]{\includegraphics[scale=0.375]{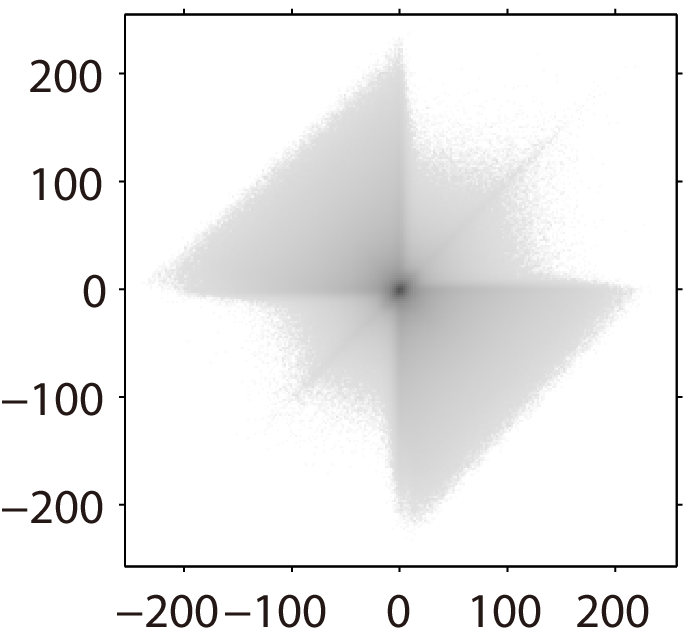}}\\
  \includegraphics[scale=0.45]{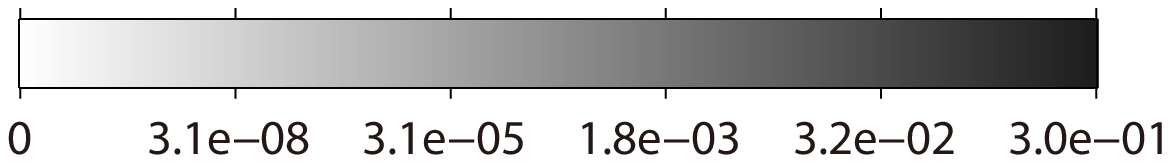}
  \caption{The average joint probability $P(x,y)$. (a) Original image. (b) Gamma correction. (c) Histogram equalization. (d) Sharpening. (e) Mean filtering. (f) Gaussian filtering. (g) Wiener filtering. (h) Median filtering. (i) Scaling. (j) Rotation. (k) JPEG compression. (l) JPEG 2000 compression. (m)(n) Anti-forensics of JPEG compression with ``dither'' and ``dither \& deblocking'', respectively. (o)(p) Anti-forensics of contrast enhancement with Cao's and Kwok's methods, respectively. (q)(r)(s) Anti-forensics of resizing with ``median'', ``edge'', ``dual''\protect\footnotemark, respectively. (t)(u)(v) Anti-forensics of rotation with  ``median'', ``edge'', ``dual'', respectively. (w) Anti-forensics of median filtering. Refer to Table \ref{table:parameter} for the details of each operations.}
  \label{fig:stat}
\end{figure*}

For an image $\bf I$, let ${\bf I}(i,j)$ be the pixel value in the $i$th row and the $j$th column. Then its horizontal backward and forward differences can be respectively written as
$$d_b(i,j)={\bf I}(i,j)-{\bf I}(i,j-1)$$
$$d_f(i,j)={\bf I}(i,j)-{\bf I}(i,j+1)$$
where $d_b(i,j), d_b(i,j) \in \{-255,\ldots,-1,0,1,\ldots,255\}$.

\footnotetext{Three anti-forensic operations were proposed in \cite{Kirchner2008}, where the first one (denoted as ``median'') was based on median filtering, and the second one (denoted as ``edge'') was based on geometric distortion with edge modulation, while the last one (denoted as ``dual'') was a dual path approach that consisted of the first two operations.}

\begin{figure*}[t]
\centering
\includegraphics[height=4cm]{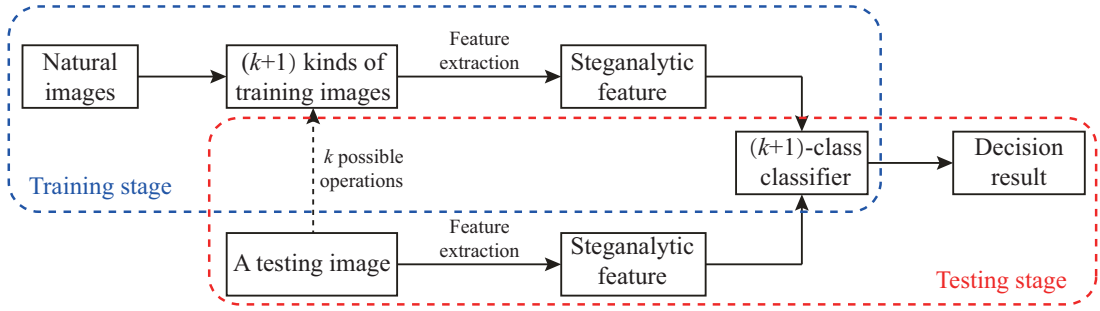}
\caption{The diagram of proposed strategy} \label{fig:procedures}
\end{figure*}

Please note that the above two differences have two important properties. First, if a pixel ${\bf I}(i,j)$ locates at smooth areas, the corresponding $d_b(i,j)$ and $d_f(i,j)$ are usually with small values, while $d_b(i,j)$ and $d_f(i,j)$ become large for the pixels located at textural areas. Second, a pixel ${\bf I}(i,j)$ with $d_f(i,j)\times d_f(i,j)>0$ is a local minimum/maximum pixel in the horizontal direction, which can be regarded as an impulse noise. In order to show the correlations among the pixel ${\bf I}(i,j)$ and its adjacent pixels, we consider the joint probability of horizontal backward and forward differences as follows
$$P(x,y) = {\bf Pr}(d_b(i,j)=x,d_f(i,j)=y)$$
where $x,y\in\{-255,\ldots,-1,0,1,\ldots,255\}$.

In Fig. \ref{fig:stat}, we illustrate the average joint probability $P(x,y)$ for 10,000 original images from BossBase v1.01 and their modified versions with various image operations. It is observed from Fig. \ref{fig:stat} that the ``shape'' of joint probability for original images (i.e. Fig. \ref{fig:stat}(a)) would become more or less different after various image operations (see Fig. \ref{fig:stat}(b)-(w)), indicating that there are some changes in the correlation measured by the joint probability $P(x,y)$. Furthermore, the ``shapes'' of joint probability for some image operations also differ from each other. Specifically, we can clearly observe that:
\begin{itemize}
\item For some spatial enhancement operations, such as histogram equalization (i.e. Fig. \ref{fig:stat}(c)) and sharpening (i.e. Fig. \ref{fig:stat}(d)), they will obviously enhance the high frequency components in an image, resulting in the values of $P(x,y)$ increase for large $(x,y)$. Similarly, Gamma correction (i.e. Fig. \ref{fig:stat}(b)) and its corresponding anti-forensic operations (i.e. Fig. \ref{fig:stat}(o) and (p)) also enlarge the values of $P(x,y)$ located at large $(x,y)$, though the degrees are less than those of histogram equalization and sharpening.

  \vspace{0.5em}
  \item Considering some spatial filtering with low-pass property, such as mean filtering (i.e. Fig. \ref{fig:stat}(e)), Gaussian filtering (i.e. Fig. \ref{fig:stat}(f)), and Wiener filtering (i.e. Fig. \ref{fig:stat}(g)), the values of $P(x,y)$ for large $(x,y)$ (which represent high frequency components) become relatively smaller, for the reason that the filtering operations tend to remove the noises and smooth the textural regions within an image. Especially, it is observed that the $P(x,y)$ with non-zero values are highly concentrated in the center (i.e. $(x,y)=(0,0)$) in Fig. \ref{fig:stat}(e) and Fig. \ref{fig:stat}(f).

  \vspace{0.5em}
  \item In Fig. \ref{fig:stat}(h), (n), (q), (t) and (w), the values of $P(x,y)$ in the 1st quadrant and the 3rd quadrant become much smaller compared to the original one (i.e. Fig. \ref{fig:stat}(a)), indicating that most local minimum/maximum pixels are removed after applying the corresponding operations. The reason is that all the corresponding operations involve median filtering, which would significantly suppress the impulse noises in an original image.

  \vspace{0.5em}
  \item The image resampling operations including scaling and rotation need to use interpolation to produce new pixels. Due to the low-pass nature of interpolation, the high frequency components within the original image are expected to be suppressed. It can be observed in Fig. \ref{fig:stat} (i) and (j) that the values of $P(x,y)$ for large $(x,y)$ indeed decrease. Moreover, it is noted that the shapes of Fig. \ref{fig:stat}(i) and (j) differ from those of Fig. \ref{fig:stat} (e), (f) and (g), though both resampling and spatial filtering mentioned above have  low-pass properties.
\end{itemize}

Overall, the above observations have validated that image operations would inevitably change the joint probability of difference between two adjacent pixels in original images, and their manners and/or degrees are usually different. Therefore, the resulting joint probabilities for some image operations become quite different from each other, which means that even such a simple feature may be helpful for exposing original images from the modified ones, and it can be further used for identifying some types of image operations.

Please note that the advanced steganalytic features such as SPAM \cite{Pevny2010}, SRM \cite{Fridrich2012}, and LBP (local binary pattern) \cite{Shi2012} applied more complicated features for better measuring the inherent correlations among adjacent pixels in original images, and they usually achieve much better detection performances. However, these features are too sophisticated and high dimensional to show them intuitively in a figure. Thus we just illustrate the joint probability of two adjacent differences in the horizontal direction for display purpose. Based on above observations and analysis, it is expected that advanced steganalytic features are very promising for detecting various image operations.

\section{The Proposed Strategy}
\label{Sec:Proposed}

In this section, we will describe the implementation of the proposed strategy. The proposed strategy consists of training and testing stages as illustrated in Fig. \ref{fig:procedures}. Compared to conventional forensic methods, the main difference in the proposed strategy is to analyze and introduce some proper steganalytic features in image forensics rather than considering some special artifacts like that in existing methods. Besides, the multiple classification forensic problem is firstly considered, i.e. the case of $k>1$.

\subsection{Feature Selection and Analysis}
\label{subsec:feature_selection}

\begin{table*}[t]
\centering
\caption{The types of image processing operations and their parameters used in the experiments.}\label{table:parameter}
\begin{tabular}{|c|l|l|} \hline
  \multicolumn{2}{|c|}{Type} & \multicolumn{1}{c|}{Parameters} \\ \hline \hline
  \multirow{2}{*}{Spatial}
  & Gamma Correction  (GC)    & $\gamma$: $0.5, \ 0.6, \ 0.7, \ 0.8, \ 0.9, \ 1.2, \ 1.4, \ 1.6, \ 1.8, \ 2.0$ \\ \cline{2-3}
  \multirow{2}{*}{enhancement}
  & Histogram Equalization  (HE) & - \\ \cline{2-3}
  & Unsharp Masking sharpening (UM)    & $\sigma$: $0.5 \sim 1.5$, $\lambda$: $0.5 \sim 1.5$ \\ \hline
  & Mean Filtering  (MeanF)  & hsize: $3\times 3, \ 5\times 5, \ 7\times 7$ \\ \cline{2-3}
  Spatial  & Gaussian Filtering  (GF) & hsize: $3\times 3, \ 5\times 5, \ 7\times 7$; $\sigma$: $0.8 \sim 1.6$  \\ \cline{2-3}
  filtering& Median Filtering  (MedF)   & hsize: $3\times 3, \ 5\times 5, \ 7\times 7$  \\ \cline{2-3}
  & Wiener Filtering  (WF)   & hsize: $3\times 3, \ 5\times 5, \ 7\times 7$ \\ \hline
  \multirow{3}{*}{Resampling}
  &\multirow{2}{*}{Scaling  (Sca)}& up-sampling: $1, \ 3, \ 5, \ 10, \ 20, \ 30, \ 40, \ 50, \ 60, \ 70, \ 80, \ 90$ (\%) \\
  &                             & down-sampling: $1, \ 3, \ 5, \ 10, \ 15, \ 20, \ 25, \ 30, \ 35, \ 40, \ 45$ (\%) \\ \cline{2-3}
  & Rotation  (Rot)           & degree:\ $1, \ 3, \ 5, \ 10, \ 15, \ 20, \ 25, \ 30, \ 35, \ 40, \ 45$ ($^\circ$) \\ \hline
  Lossy & JPEG               & quality factor: $75 \sim 99$ \\ \cline{2-3}
  compression & JPEG 2000  (JP2)   & compression ratio: $ 2.0 \sim 8.0$ \\ \hline
\end{tabular}
\end{table*}

It is well known that feature selection plays a very important role in various classification algorithms. Up to now, lots of steganalytic methods have been proposed. It is known that the steganalytic methods can be divided into two different types, \emph{i.e.} targeted and universal methods \cite{Li2011}. Targeted steganalytic methods are mainly based on some special artifacts introduced by the targeted steganography. For example, based on the structural asymmetry of LSB (least significant bit) replacement algorithm (which never decreases even pixels and increases odd pixels), the Chi-squared attack \cite{Chi_squared} and regular/singular groups analysis \cite{RS01} can effectively detect LSB replacement even at a low embedding rate, e.g. less than 0.05 bpp. However, these methods will totally fail to detect other steganography (as well as other image operations) such as LSB matching which does not introduce the structural asymmetry at all. Therefore, targeted steganalytic features are not considered in the proposed strategy.  Unlike the targeted steganalytic methods, universal steganalytic methods (such as SRM \cite{Fridrich2012} and LBP \cite{Shi2012}) try to model some inherent statistical properties within natural images. Usually, these universal steganalytic features are generated by computing the statistics on a set of residuals of the given image, while the residuals are created by filtering the image with different masks which cover a pixel and its several neighbors. It is noted that different filter masks try to catch different types of correlations among adjacent pixels, hence the resulting features can represent different statistical properties. Once some pixels within an original image are modified, such features would changed inevitably. Therefore, the universal steganalytic features can be adopted in the proposed strategy.

\begin{table*}[t]
\centering
\caption{Average detection accuracies(\%) for identifying the original images and the images after a given type of image processing operation. The best results in each case are marked with bold font, and the underlined results denote the accuracies using the specific methods to detect the corresponding operation.}\label{table:identify}
\begin{tabular}{|c|c|c|c|c|c|c|c|c|c|c|c| } \hline
          & GC    & HE    & UM    & MeanF    & GF    & MedF    & WF    & Sca   & Rot   & JPEG  & JP2   \\ \hline \hline
  CE      & \textcolor{blue}{\underline{97.67}} & \textcolor{blue}{\underline{99.52}} & 55.39 & 65.14 & 62.17 & 78.96 & 65.63 & 56.05 & 54.56 & 54.78 & 53.33 \\ \hline
  CEBF    & \textcolor{blue}{\underline{95.25}} & \textcolor{blue}{\underline{98.70}} & 50.07 & 50.16 & 50.06 & 52.81 & 50.04 & 50.08 & 50.04 & 50.17 & 50.05 \\ \hline
  AR      & 52.67 & 64.74 & 64.25 & 93.87 & 91.53 & \textcolor{blue}{\underline{90.68}} & 72.41 & 69.17 & 77.69 & 61.80 & 57.92 \\ \hline
  GLF     & 89.32 & 99.55 & 93.58 & 99.98 & 99.97 & \textcolor{blue}{\underline{99.98}} & 99.88 & 81.18 & 93.93 & 90.65 & 93.74 \\ \hline
  PPI     & 50.48 & 53.04 & 50.13 & 52.71 & 52.64 & 57.69 & 61.94 & \textcolor{blue}{\underline{91.34}} & \textcolor{blue}{\underline{90.86}} & 84.46 & 85.18 \\ \hline
  JPA     & 54.73 & 67.68 & 71.65 & 97.03 & 97.80 & 88.18 & 95.26 & 69.83 & 64.61 & \textcolor{blue}{\underline{96.25}} & 70.40 \\ \hline
  Combined& \textbf{98.42} & 99.96 & 93.96 & 99.98 & 99.98 & 99.98 & 99.91 & 93.09 & 97.55 & \textbf{97.82} & 93.76 \\ \hline \hline

  SRM     & 98.05 & \textbf{99.97}   & \textbf{99.35} & \textbf{100}   & \textbf{100}   & \textbf{100}   & \textbf{100}   & \textbf{97.99} & \textbf{99.95} & 97.78 & \textbf{99.95} \\ \hline
  LBP     & 90.96 & 99.42 & 97.60 & 99.99 & 99.99 & \textbf{100}   & 99.99 & 96.33 & \textbf{99.95} & 97.76 & 99.53 \\ \hline \hline

  Combined without PPI& 98.40 & 99.94 & 93.98 & 99.98 & 99.98 & 99.96 & 99.91 & \textcolor{red}{82.19} & \textcolor{red}{94.58} & 97.79 & 93.78 \\ \hline
\end{tabular}
\end{table*}

\subsection{The Methodology of Classification}
\label{subsec:classification}
After the features are selected, we can perform classification with the supervised learning  scheme as illustrated in Fig. \ref{fig:procedures}. Assume that the number of possible image processing and/or anti-forensic operations is $k$ ($k\geq1$) for a given testing image, we need to design a ($k+1$)-class (including the class of original images) classifier. Please note that when $k=1$, it is a similar case with the existing forensic works, i.e., determining whether a given image has been modified with a specific operation. Thus a binary classifier is needed in such a case. When $k>1$, we adopt the pairwise coupling strategy \cite{Knerr1990} to train a multi-class classifier. Specifically, for all the (k+1) classes labeled with integers $\{1,2,\ldots,k+1\}$, we select each possible pair of class $i$ and class $j$ ($i,j\in\{1,2,\ldots,k+1\}$, and $i\neq j$) to train a binary classifier respectively, and totally obtain $N=k\times (k+1)/2$ classifiers, each of which outputs a label either be $i$ or $j$ for an input feature. For a testing image, hence, we feed its features to all the $N$ classifiers and obtain $N$ class labels. Finally we choose the most frequently occurring label as the predicted class label of this testing image.

\section{Experimental Results}
\label{Sec:Results}

In the experiments, 10,000 raw images were downloaded from the Boss Base v1.01 \cite{Bas2011}. These images were firstly converted into gray scale bitmaps, and then a center region with size of $1024 \times 1024$ was cropped for each image, and finally like the pre-operations in \cite{Kirchner2008,Popescu2005b}, all selected segments were down-sampled with a factor 2 to remove the artifacts of demosaicing introduced by CFA (color filter array) interpolation within digital cameras \cite{Swaminathan2007}. Therefore, all the resulting images with size of $512 \times 512$ were regarded as original images. Please note that just gray scale images are considered in our experiments in this paper. We can obtain similar results for color images when performing the proposed strategy on the illumination channel of color images based on our experiments.

In each of the following experiments, we used the ensemble classifier \cite{Kodovsky2012} with its default settings for classification. All the images were randomly divided into two categories: 50\% was used for training and the rest for testing. We repeat the training and testing 10 times and show the average results in subsection \ref{subsec:postprocess_detection} and \ref{subsec:Counter AF}.

\subsection{Exposing image processing operations}
\label{subsec:postprocess_detection}

In this subsection, we use the proposed strategy to detect various image processing operations. Both binary classification (see 1)) and multi-class classification (see 2)) are considered.

\begin{table*}[t]
\centering
\caption{Confusion matrix for identifying the types of operations using the SRM features \cite{Fridrich2012}. Please note that the asterisk ``$\ast$'' here  denotes that the corresponding accuracy is less than 1\%. }\label{table:mulSRM}
\begin{tabular}{|c||c|c|c|c|c|c|c|c|c|c|c|c|}\hline
  Actual$\setminus$Predicted & Orig & GC & HE & UM & MeanF & GF & MedF & WF & Sca & Rot & JPEG & JP2 \\ \hline \hline
  Orig & \textbf{97.57}  & $\ast$ & $\ast$ & $\ast$ & $\ast$ & $\ast$ & $\ast$ & $\ast$ & 1.05  & $\ast$ & $\ast$ & $\ast$ \\ \hline
  GC & 2.86  & \textbf{96.37}  & $\ast$ & $\ast$ & $\ast$ & $\ast$ & $\ast$ & $\ast$ & $\ast$ & $\ast$ & $\ast$ & $\ast$ \\ \hline
  HE & $\ast$ & $\ast$ & \textbf{99.56}  & $\ast$ & $\ast$ & $\ast$ & $\ast$ & $\ast$ & $\ast$ & $\ast$ & $\ast$ & $\ast$ \\ \hline
  UM & $\ast$ & $\ast$ & $\ast$ & \textbf{99.11}  & $\ast$ & $\ast$ & $\ast$ & $\ast$ & $\ast$ & $\ast$ & $\ast$ & $\ast$ \\ \hline
  MeanF & $\ast$ & $\ast$ & $\ast$ & $\ast$ & \textbf{99.23}  & $\ast$ & $\ast$ & $\ast$ & $\ast$ & $\ast$ & $\ast$ & $\ast$ \\ \hline
  GF & $\ast$ & $\ast$ & $\ast$ & $\ast$ & $\ast$ & \textbf{99.69}  & $\ast$ & $\ast$ & $\ast$ & $\ast$ & $\ast$ & $\ast$ \\ \hline
  MedF & $\ast$ & $\ast$ & $\ast$ & $\ast$ & $\ast$ & $\ast$ & \textbf{99.90}  & $\ast$ & $\ast$ & $\ast$ & $\ast$ & $\ast$ \\ \hline
  WF & $\ast$ & $\ast$ & $\ast$ & $\ast$ & 1.39  & $\ast$ & $\ast$ & \textbf{98.53}  & $\ast$ & $\ast$ & $\ast$ & $\ast$ \\ \hline
  Sca & 3.13  & $\ast$ & $\ast$ & $\ast$ & $\ast$ & $\ast$ & $\ast$ & $\ast$ & \textbf{96.17}  & $\ast$ & $\ast$ & $\ast$ \\ \hline
  Rot & $\ast$ & $\ast$ & $\ast$ & $\ast$ & $\ast$ & $\ast$ & $\ast$ & $\ast$ & $\ast$ & \textbf{99.25}  & $\ast$ & $\ast$ \\ \hline
  JPEG & 4.04  & $\ast$ & $\ast$ & $\ast$ & $\ast$ & $\ast$ & $\ast$ & $\ast$ & $\ast$ & $\ast$ & \textbf{95.67}  & $\ast$ \\ \hline
  JP2 & $\ast$ & $\ast$ & $\ast$ & $\ast$ & $\ast$ & $\ast$ & $\ast$ & $\ast$ & $\ast$ & $\ast$ & $\ast$ & \textbf{99.84}  \\ \hline
\end{tabular}
\end{table*}

\begin{table*}[t]
\centering
\caption{Confusion matrix for identifying the types of operations using the LBP features \cite{Shi2012}. Please note that the asterisk ``$\ast$'' here  denotes that the corresponding accuracy is less than 1\%. }\label{table:mulLBP}
\begin{tabular}{|c||c|c|c|c|c|c|c|c|c|c|c|c|}\hline
  Actual$\setminus$Predicted & Orig & GC & HE & UM & MeanF & GF & MedF & WF & Sca & Rot & JPEG & JP2 \\ \hline \hline
Orig & \textbf{92.84}  & 3.19  & $\ast$ & 1.93  & $\ast$ & $\ast$ & $\ast$ & $\ast$ & 1.51  & $\ast$ & $\ast$ & $\ast$ \\ \hline
GC & 13.56  & \textbf{83.14}  & 1.19  & 1.40  & $\ast$ & $\ast$ & $\ast$ & $\ast$ & $\ast$ & $\ast$ & $\ast$ & $\ast$ \\ \hline
HE & $\ast$ & 1.33  & \textbf{98.44}  & $\ast$ & $\ast$ & $\ast$ & $\ast$ & $\ast$ & $\ast$ & $\ast$ & $\ast$ & $\ast$ \\ \hline
UM & 2.12  & $\ast$ & $\ast$ & \textbf{96.60}  & $\ast$ & $\ast$ & $\ast$ & $\ast$ & $\ast$ & $\ast$ & $\ast$ & $\ast$ \\ \hline
MeanF & $\ast$ & $\ast$ & $\ast$ & $\ast$ & \textbf{99.25}  & $\ast$ & $\ast$ & $\ast$ & $\ast$ & $\ast$ & $\ast$ & $\ast$ \\ \hline
GF & $\ast$ & $\ast$ & $\ast$ & $\ast$ & $\ast$ & \textbf{99.75}  & $\ast$ & $\ast$ & $\ast$ & $\ast$ & $\ast$ & $\ast$ \\ \hline
MedF & $\ast$ & $\ast$ & $\ast$ & $\ast$ & $\ast$ & $\ast$ & \textbf{99.75}  & $\ast$ & $\ast$ & $\ast$ & $\ast$ & $\ast$ \\ \hline
WF & $\ast$ & $\ast$ & $\ast$ & $\ast$ & 1.74  & $\ast$ & $\ast$ & \textbf{98.13}  & $\ast$ & $\ast$ & $\ast$ & $\ast$ \\ \hline
Sca & 5.30  & $\ast$ & $\ast$ & $\ast$ & $\ast$ & $\ast$ & $\ast$ & $\ast$ & \textbf{93.20}  & $\ast$ & $\ast$ & $\ast$ \\ \hline
Rot & $\ast$ & $\ast$ & $\ast$ & $\ast$ & $\ast$ & $\ast$ & $\ast$ & $\ast$ & 1.50  & \textbf{98.42}  & $\ast$ & $\ast$ \\ \hline
JPEG & 3.78  & $\ast$ & $\ast$ & $\ast$ & $\ast$ & $\ast$ & $\ast$ & $\ast$ & $\ast$ & $\ast$ & \textbf{95.60}  & $\ast$ \\ \hline
JP2 & $\ast$ & $\ast$ & $\ast$ & $\ast$ & $\ast$ & $\ast$ & $\ast$ & $\ast$ & $\ast$ & $\ast$ & $\ast$ & \textbf{99.03}  \\ \hline
\end{tabular}
\end{table*}

\begin{table*}[t]
\centering
\caption{Confusion matrix for identifying the types of operations using the six combined features. Please note that the asterisk ``$\ast$'' here  denotes that the corresponding accuracy is less than 1\%. }\label{table:mulCOM}
\begin{tabular}{|c||c|c|c|c|c|c|c|c|c|c|c|c|}\hline
  Actual$\setminus$Predicted & Orig & GC & HE & UM & MeanF & GF & MedF & WF & Sca & Rot & JPEG & JP2 \\ \hline \hline
  Orig & \textbf{86.45}  & $\ast$ & $\ast$ & 5.54  & $\ast$ & $\ast$ & $\ast$ & $\ast$ & 4.54  & 1.47  & $\ast$ & 1.00  \\ \hline
  GC & 2.06  & \textbf{93.86}  & 2.17  & $\ast$ & $\ast$ & $\ast$ & $\ast$ & $\ast$ & $\ast$ & $\ast$ & $\ast$ & $\ast$ \\ \hline
  HE & $\ast$ & 2.05  & \textbf{97.95}  & $\ast$ & $\ast$ & $\ast$ & $\ast$ & $\ast$ & $\ast$ & $\ast$ & $\ast$ & $\ast$ \\ \hline
  UM & 5.95  & $\ast$ & $\ast$ & \textbf{92.80}  & $\ast$ & $\ast$ & $\ast$ & $\ast$ & $\ast$ & $\ast$ & $\ast$ & $\ast$ \\ \hline
  MeanF & $\ast$ & $\ast$ & $\ast$ & $\ast$ & \textbf{88.94}  & 6.84  & $\ast$ & 3.23  & $\ast$ & $\ast$ & $\ast$ & $\ast$ \\ \hline
  GF & $\ast$ & $\ast$ & $\ast$ & $\ast$ & 6.83  & \textbf{90.91}  & $\ast$ & 1.12  & $\ast$ & $\ast$ & $\ast$ & $\ast$ \\ \hline
  MedF & $\ast$ & $\ast$ & $\ast$ & $\ast$ & $\ast$ & $\ast$ & \textbf{97.18}  & 1.83  & $\ast$ & $\ast$ & $\ast$ & $\ast$ \\ \hline
  WF & $\ast$ & $\ast$ & $\ast$ & $\ast$ & 3.75  & 1.22  & 1.87  & \textbf{91.37}  & $\ast$ & $\ast$ & $\ast$ & 1.20  \\ \hline
  Sca & 7.54  & $\ast$ & $\ast$ & $\ast$ & $\ast$ & $\ast$ & $\ast$ & $\ast$ & \textbf{76.16}  & 13.49  & $\ast$ & $\ast$ \\ \hline
  Rot & 2.25  & $\ast$ & $\ast$ & $\ast$ & $\ast$ & $\ast$ & $\ast$ & $\ast$ & 16.99  & \textbf{79.59}  & $\ast$ & $\ast$ \\ \hline
  JPEG & 3.67  & $\ast$ & $\ast$ & $\ast$ & $\ast$ & $\ast$ & $\ast$ & $\ast$ & $\ast$ & $\ast$ & \textbf{94.59}  & $\ast$ \\ \hline
  JP2 & 10.72  & $\ast$ & $\ast$ & $\ast$ & $\ast$ & $\ast$ & $\ast$ & $\ast$ & 1.76  & $\ast$ & $\ast$ & \textbf{84.48}  \\ \hline
\end{tabular}
\end{table*}

\vspace{0.5em}
\subsubsection{Detection of a single operation}
In this subsection, we try to determine whether or not a questionable image has been previously subject to a given image processing operation, including spatial enhancement, lossy compression, filtering and so on. The parameters of these operations are shown in Table \ref{table:parameter}.

For each original image, we created 11 counterparts using a random parameter selected in Table \ref{table:parameter} for each operation. Two steganalytic features are employed in the proposed strategy, \emph{i.e.}, SRM \cite{Fridrich2012} and LBP \cite{Shi2012}\footnote{We use these two feature sets with their default parameters as introduced in \cite{Fridrich2012} and \cite{Shi2012}, so the dimensions of them are 34671 and 22153, respectively.}. Besides, six state-of-the-art forensic features including CE \cite{Stamm2008}, CEBF \cite{Cao2014}, AR \cite{Kang2012}, GLF \cite{Chen2013a},  PPI \cite{Mahdian2008}, JPA \cite{Luo2010} and their combined version (denote as ``Combined'') were included for comparative studies. For each operation, therefore, we trained totally 9 classifiers corresponding to 9 different feature sets, where each classifier was trained with the feature vectors extracted from the manipulated images and the corresponding original images in the training category. Finally, we used the trained classifiers to predict class labels for the testing images and obtained the detection accuracies.

\begin{table*}[t]
\centering
\caption{Average accuracies (\%) along the diagonal direction in the corresponding confusion matrix.}\label{table:judge}
\begin{tabular}{|c|c|c|c|c|c|c|c||c|c|} \hline
  Features & ~CE~    & CEBF  & ~AR~    & GLF~   & PPI~   & JPA~   & Combined & SRM~ & LBP~ \\ \hline
  Accuracy & 27.00 & 17.78 & 35.84 & 79.91 & 19.60 & 30.40 & 89.52 & 98.41 & 96.18 \\ \hline
\end{tabular}
\end{table*}

\begin{table*}[t]
  \centering
  \caption{The classification accuracies for detecting anti-forensics of JPEG compression.}
  \begin{tabular}{|c|c||c|c|c|c|c|c|}
  \hline
  \multicolumn{2}{|c||}{QF} & 75 & 80 & 85 & 90 & 95 & Random \\ \hline \hline
  \multirow{5}{*}{dither}
  & Method \cite{Valenzise2011} & 88.83 & 87.81 & 86.47 & 84.79 & 58.50 & 74.46 \\ \cline{2-8}
  & 1st detector \cite{Lai2011} & 83.38 & 80.74 & 76.49 & 68.59 & 55.64 & 69.76 \\ \cline{2-8}
  & 2nd detector \cite{Lai2011} & 98.89 & 97.94 & 93.75 & 91.91 & 85.70 & 89.55 \\ \cline{2-8}
  & SRM & \textbf{99.95} & \textbf{99.90} & \textbf{99.85} & \textbf{99.77} & \textbf{99.34} & \textbf{96.38} \\ \cline{2-8}
  & SPAM   & \textbf{97.92} & \textbf{97.28} & \textbf{96.30} & \textbf{94.51} & \textbf{90.37} & \textbf{88.50} \\ \hline\hline
  \multirow{2}{*}{dither}
  & Method \cite{Valenzise2011} & 49.99 & 50.00 & 49.99 & 49.99 & 50.00 & 50.00 \\ \cline{2-8}
  & 1st detector \cite{Lai2011} & 60.08 & 60.42 & 60.87 & 61.10 & 61.47 & 60.73 \\ \cline{2-8}
  \& & 2nd detector \cite{Lai2011} & 76.13 & 71.99 & 67.76 & 62.87 & 54.76 & 63.72 \\ \cline{2-8}
  \multirow{2}{*}{deblocking}
  & SRM & \textbf{99.99} & \textbf{99.98} & \textbf{99.99} & \textbf{99.98} & \textbf{99.99} & \textbf{99.97} \\ \cline{2-8}
  & SPAM  & \textbf{99.81} & \textbf{99.78} & \textbf{99.78} & \textbf{99.77} & \textbf{99.77} & \textbf{99.78} \\ \hline
  \end{tabular}
  \label{tab:rst_JPEG}
\end{table*}

The average detection accuracies are shown in Table \ref{table:identify}. It is observed that SRM \cite{Fridrich2012} and LBP \cite{Shi2012} usually perform the best or nearly the best in all cases, indicating that both the steganalytic features are very useful for detecting different operations. For the six specific forensic methods, although their detection performances for the corresponding operations are good (see the underlined values in Table \ref{table:identify}), their performances  are rather poor for other operations. For instance, the method AR \cite{Kang2012} can effectively detect the median filtering, Gaussian filtering and mean filtering with accuracies larger than 90\%, while it fails to detect Gamma correction and the corresponding accuracy drops to 52.67\%, which is very close to the random guessing. Please note that the combined features provide quite satisfactory results in this experiment.
However, such combined features are still regarded as targeted features, since when an image with a new operation is tested, the corresponding features about the new operation has to be carefully designed and added. Otherwise, the performance for the new operation would be very poor. To validate this issue, we have tested another set of combined features which consists of five aforementioned forensic features without PPI (the feature for resampling detection). We show the detection results in the last row of Table \ref{table:identify}. As expected, the accuracies for detecting resampling become obviously poor compared to that of the combined version with six features, especially for scaling, the decrement is over 10\% in this case. It is also noted that the accuracies for detecting other operations are nearly unchanged, since only the PPI feature is excluded. In contrast, the steganalytic features would not change  at all even when new image operations are considered.

\vspace{0.5em}
\subsubsection{Identifying the type of various operations}
\label{subsubsec:multiclass}

In this subsection, we try to identify the type of several possible image processing previously used for a given questionable image. All types of operations listed in Table  \ref{table:parameter} are considered in this experiment, \emph{i.e.}, this experiment involves a 12-class classification problem. The test images were created similarly as described in subsection \ref{subsec:postprocess_detection} using the parameters randomly selected in Table \ref{table:parameter}.

We also evaluated the proposed scheme with SRM \cite{Fridrich2012} and LBP \cite{Shi2012}. The confusion matrices are shown in Table \ref{table:mulSRM} and Table \ref{table:mulLBP}, respectively. It is observed that both steganalytic features can effectively identify the type of operation for a given image, especially the SRM \cite{Fridrich2012}. On average, the detection accuracies averaging along the diagonal direction in the two confusion matrices are  98.41\% and 96.18\%, respectively. For comparative study, we show the confusion matrix for the combined features (which work the best among special features in binary classification as shown in Table \ref{table:identify}) in Table \ref{table:mulCOM}, and show the average results along the diagonals of confusion matrices for all the tested features in Table \ref{table:judge}. From the two tables, it is observed that the detection performances of most targeted methods are rather poor. Even the combined features can obtain satisfactory results, it still obviously poorer than ours with the average accuracy lower than 90\%.

\subsection{Exposing anti-forensic operations}
\label{subsec:Counter AF}

In this subsection, we apply the proposed strategy to exposing four existing typical anti-forensic operations. In the following, the experimental results for detecting each operation are firstly shown, and then a discussion on the universality of the proposed strategy is given.

\vspace{0.5em}
\subsubsection{Detecting anti-forensics of JPEG compression}
\label{subsec:Anti-Single}
In this experiment, we firstly JPEG compressed the original images using six quality factors (QF) (the first five QF are ranging from 75 to 95 with a step of 5, while the remaining one is randomly selected between 75 and 99), and obtained totally six categories of JPEG images, then we decompressed them into the spatial domain. Finally we performed the two anti-forensic operations described in \cite{Stamm2011a} (namely, adding dithers to DCT coefficients and reducing the blocking artifacts after addition of dithers, denoted as ``dither'' and ``dither \& deblocking'' for short) to obtain the resulting bitmap images as positive instances, while the negative instances were the original uncompressed images.

We adopted SRM and SPAM to differentiate the anti-forensically modified images from the original ones. For comparative studies, the existing countering anti-forensic methods \cite{Valenzise2011,Lai2011} were included. The experimental results are shown in Table \ref{tab:rst_JPEG}. Obviously, it is observed that the steganalytic features can achieve very good performances with average accuracy rates over 96\%, which is much better than all existing countering methods, especially for detecting ``dither \& deblocking''. Note that the method proposed in \cite{Valenzise2011} becomes useless if removal of blocking artifacts is applied, since it only tries to reveal the traces introduced by anti-forensic dithers.

\begin{table}[t]
  \centering
  \caption{The classification accuracies for detecting anti-forensics of contrast enhancement}
  \begin{tabular}{|c|c||c|c|c|c|c|} \hline
    \multicolumn{2}{|c||}{$\gamma$}   &  0.6  &  0.8  &  1.2  &  1.6  & Random \\ \hline \hline
    \multirow{2}{*}{Method \cite{Cao2010}} & SRM  & 91.36 & 89.41 & 87.87 & 90.69 & 86.72 \\ \cline{2-7}
    & SPAM & 86.63 & 84.37 & 82.84 & 86.63 & 82.19 \\ \hline \hline
    \multirow{2}{*}{Method \cite{Kwok2011}} & SRM & 87.75 & 79.15 & 73.53 & 87.82 & 72.99 \\ \cline{2-7}
    & SPAM & 84.15 & 75.10 & 70.44 & 84.93 & 69.83 \\ \hline
  \end{tabular}
  \label{tab:rst_cont}
\end{table}

\begin{table}[t]
  \centering
  \caption{The classification accuracies for detecting anti-forensics of resampling \#1: Scaling}{
  \begin{tabular}{|c|c||c|c|c|c|c|} \hline
    \multicolumn{2}{|c||}{$s$}
    &  0.6  &  0.8  &  1.2  &  1.6  &  Random \\ \hline \hline
    \multirow{2}{*}{Median}
    & SRM  &  100  &  99.99  &  100  &  100  &  100 \\ \cline{2-7}
    & SPAM &  100  &  99.99  &  99.99  &  99.99  &  99.99 \\ \hline \hline
    \multirow{2}{*}{Edge}
    & SRM  &  99.82  &  99.98  &  99.99  &  100  &  99.71 \\ \cline{2-7}
    & SPAM &  96.58  &  99.67  &  99.92  &  99.97  &  97.64 \\ \hline \hline
    \multirow{2}{*}{Dual}
    & SRM  &  99.95  &  99.97  &  99.99  &  99.99  &  99.95 \\ \cline{2-7}
    & SPAM &  98.36  &  99.25  &  99.93  &  99.97  &  98.68 \\ \hline
  \end{tabular}}
  \label{tab:rst_resizing}
\end{table}

\begin{table}[t]
  \centering
  \caption{The classification accuracies for detecting anti-forensics of resampling \#2: Rotation}{
  \begin{tabular}{|c|c||c|c|c|c|c|c|c|c|c|} \hline
    \multicolumn{2}{|c||}{$\theta$}
    & $10^\circ$ & $20^\circ$ & $30^\circ$ & $40^\circ$ & Random \\ \hline \hline
    \multirow{2}{*}{Median}
    & SRM  &  100   &  99.99  &  99.99  &  99.99  &  99.99 \\ \cline{2-7}
    & SPAM &  99.99   &  100  &  99.99  &  100  &  99.99 \\ \hline \hline
    \multirow{2}{*}{Edge}
    & SRM  &  99.99 &  99.99  &  99.98  &  99.99  &  99.99 \\ \cline{2-7}
    & SPAM &  99.91 &  99.88  &  99.87  &  99.83  &  99.87 \\ \hline \hline
    \multirow{2}{*}{Dual}
    & SRM  &  99.97 &  99.98  &  99.99  &  99.98  &  99.97 \\ \cline{2-7}
    & SPAM &  99.83 &  99.87  &  99.88  &  99.86  &  99.85 \\ \hline
  \end{tabular}}
  \label{tab:rst_rotating}
\end{table}

\begin{table}[t]
  \centering
  \caption{The classification accuracies for detecting anti-forensics of median filtering.}{
  \begin{tabular}{|c||c|c|c|c|c|}
  \hline
  \multirow{2}{*}{} & \multicolumn{2}{c|}{$B=4$} & \multicolumn{2}{c|}{$B=8$} & \multirow{2}{*}{Random} \\ \cline{2-5}
  & $T=2$ & $T=4$ & $T=2$ & $T=4$ & \\ \hline \hline
  Method \cite{Zeng2014} & 99.91 & 99.90 & 99.97 & 99.95 & 99.91  \\ \hline \hline
  SRM & \textbf{100} & \textbf{99.99} & \textbf{100} & \textbf{99.99} & \textbf{100}   \\ \hline
  SPAM & \textbf{99.95} & \textbf{99.89} & \textbf{99.92} & \textbf{99.87} & \textbf{99.84}  \\ \hline

  \end{tabular}}
  \label{tab:rst_median}
\end{table}

\vspace{0.5em}
\subsubsection{Detecting anti-forensics of contrast enhancement}
\label{subsec:Anti-enhancement}
In this experiment, we firstly enhanced the contrast of each original image via Gamma correction with 5 parameters ($\gamma = 0.6, 0.8, 1.2, 1.6$, and one randomly selected from 0.5 to 2.0 excluding 1.0), and then we respectively performed the two anti-forensic methods \cite{Cao2010} and \cite{Kwok2011} on the resulting images to obtain the test images as positive instances. In this case, the negative instances were the original images without contrast enhancement.

The experimental results are shown in Table \ref{tab:rst_cont}. Two observations can be obtained from Table \ref{tab:rst_cont}. First, compared with Cao's method \cite{Cao2010}, Kwok's method \cite{Kwok2011} is more difficult to be detected, since Kwok's method processes images like what is done inside digital camera, thus leaving fewer detectable artifacts. Second, when $\gamma$ approaches to 1, the detection performances are degraded. When the parameter $\gamma$ is selected randomly, the detection accuracies are not very satisfactory in this case, especially for Kwok's anti-forensic method. Please note that none relative work has been proposed to detect anti-forensics of contrast enhancement previously.

\begin{table*}[t]
  \centering
  \caption{The classification accuracies for detecting aforementioned anti-forensic operations. The blue texts with underlines indicate the results using the specific methods to detect their targeted operations.}{
  \begin{tabular}{|c||c|c|c|c|c|c|c|c|} \hline
    &  \multicolumn{2}{c|}{JPEG AF}        & \multicolumn{2}{c|}{CE AF}    &\multicolumn{3}{c|}{Resampling AF} & \multirow{3}{*}{MedF AF} \\ \cline{2-8}
    &  \multirow{2}{*}{dither} & dither \& &  Method  &  Method  &\multirow{2}{*}{median}&\multirow{2}{*}{edge}&\multirow{2}{*}{dual}& \\
    &                          & deblocking&  \cite{Cao2010}  &  \cite{Kwok2011}   &    &    &    &  \\ \hline \hline
  Method \cite{Valenzise2011}  & \underline{\textcolor{blue}{74.46}} & 50.00 & 50.74 & 50.00 & 69.86 & 54.02 & 50.56 & 60.81 \\ \hline
  1st detector in \cite{Lai2011} & \underline{\textcolor{blue}{69.76}} & 60.73 & 49.95 & 54.58 & 93.39 & 57.43 & 63.64 & 63.01 \\ \hline
  2nd detector in \cite{Lai2011} & \underline{\textcolor{blue}{89.55}} & 63.72 & 49.98 & 51.06 & 59.23 & 53.94 & 51.78 & 74.07 \\ \hline
  Method \cite{Zeng2014}  & 69.42 & 50.77 & 49.71 & 50.27 & 51.26 & 51.22 & 51.23 & \underline{\textcolor{blue}{99.91}} \\ \hline \hline
  SRM                     & \textbf{96.38} & \textbf{99.97} & \textbf{86.72} & \textbf{72.99} & \textbf{99.99} & \textbf{99.85} & \textbf{99.96} & \textbf{100} \\ \hline
  SPAM                    & \textbf{88.50} & \textbf{99.78} & \textbf{82.19} & \textbf{69.83} & \textbf{99.99} & \textbf{98.75} & \textbf{99.27} & \textbf{99.84} \\ \hline
  \end{tabular}}
  \label{tab:crossdetect}
\end{table*}

\vspace{0.5em}
\subsubsection{Detecting anti-forensics of resampling}
\label{subsec:Anti-Resampling}
In this experiment, two kinds of resampling operations were investigated, namely, scaling and rotation. The scaling factors used in the experiment were ranging from 0.6 to 1.8, and the rotation angles were ranging from $10^\circ$ to $40^\circ$ with a step $10^\circ$. The three anti-forensic operations \cite{Kirchner2008} (denoted as ``Median'', ``Edge'' and ``Dual'' for short) were applied to obtain the positive instances, while the corresponding negative instances were the original bitmaps without any resampling operation.

Table \ref{tab:rst_resizing} and \ref{tab:rst_rotating} show the experimental results for the resampling cases of scaling and rotation, respectively. From Table \ref{tab:rst_resizing} and \ref{tab:rst_rotating}, it is observed that the detection accuracies are over 96.58\% for all cases, achieving very good performance.

\vspace{0.5em}
\subsubsection{Detecting anti-forensics of median filtering}
In this experiment, we firstly generated 10,000 median filtered images with the filter size $3\times 3$, and then the resulting images were modified by the anti-forensic operation \cite{Wu2013} with the parameter $B=\{4,8\}$ and $T=\{2,4\}$. In this situation, the positive instances were the anti-forensically modified median filtered images, while the negative instances were the original images without median filtering. For a comparison, the recent work (Zeng \emph{et al.}'s method \cite{Zeng2014}) is  included in this experiment.

The experimental results are shown in Table \ref{tab:rst_median}. It is observed that all the detection accuracies are higher than 99.8\%, indicating that all these methods achieve very good performance. Compared with the method \cite{Zeng2014}, the proposed scheme with SRM slightly outperforms it, while the proposed scheme with SPAM can achieve almost the same performance with \cite{Zeng2014} on average.

\vspace{0.5em}
\subsubsection{Discussions on the universality}

From the results shown above, we find that the proposed strategy can effectively expose the images after the four different kinds of existing anti-forensic operations using some typical universal steganalytic features.
On the other hand, it is expected that the existing specific methods such as \cite{Valenzise2011,Lai2011,Zeng2014} would not be effective for detecting other anti-forensic operations, since their features are highly dependent on the special traces of corresponding anti-forensic operations. To verify this issue, we further use the specific methods \cite{Valenzise2011,Lai2011,Zeng2014} to detect all the aforementioned anti-forensic operations.

The experimental results are shown in Table \ref{tab:crossdetect}\footnote{In Table \ref{tab:crossdetect}, ``AF'', ``MedF'', and ``CE'' denote ``anti-forensics'', ``median filtering'', and ``contrast enhancement'', respectively. Note that the results for detecting anti-forensics of resampling are averaged on both resizing and rotation operations.}. It is observed that almost all existing specific methods fail to detect other types of anti-forensic operations with an average accurate of around 50\%. There is an occasional exceptional situation that the Lai's 1st detector \cite{Lai2011} can effectively identify anti-forensics of resampling with the ``median'' operation (over 93\%). The reason may be that the detector works by measuring the strength of high frequency components, while the ``median'' operation applies median filtering that would suppress the high frequency components, which is just match the features of Lai's 1st detector. As a comparison, it can be observed from the last two rows of Table \ref{tab:crossdetect} that the proposed scheme with SRM and SPAM features achieves very good performance in all the cases, meaning that this scheme can be universally used for countering various anti-forensic operations.

\section{Conclusion}
\label{sec:Conclusion}

In this paper, we propose a novel strategy from the view of steganalysis to detect various image operations, including various common image processing and most existing anti-forensic operations. The main contributions of the paper are as follows:

\begin{itemize}
\item We analyze the common artifacts introduced by various image operations, and show that some inherent correlation among adjacent pixels within an original image is difficult to be well preserved after any image operation, especially when the modification rate is high.

\item We analyze the similarity between image operations (including various image processing and anti-forensic operations) and steganography with detailed examples and extensive quantitative data. Then we model the image operations as data hiding and build a bridge between digital image forensics and steganalysis.

\item We adopt a strategy that applies steganalytic features for detecting various image operations. The extensive experiments show that the proposed strategy with some advanced steganalytic features significantly outperforms the existing targeted forensic methods in both effectiveness and universality. Furthermore, multi-classification is considered in our experiments.
\end{itemize}

What is more, this paper also provides some valuable insights for both forensic investigator and forger. For the investigator, it is very effective to detect various image operations with some advanced steganalytic features, and thus there is no need to just consider the special artifacts introduced by a given image operation as it did in previous forensic works. For the forgers, they should not only try to remove the tampering traces to resist the targeted detector, but also need to carefully considered some inherent statistics within the original images when performing any anti-forensic operation. Besides, the proposed strategy is flexible. With the development on steganalysis, more advanced high-dimensional steganalytic features in the future can be directly used in the proposed strategy to further improve the detection performances for exposing the existing and new image operations. Our future efforts will also be made to generalize the proposed strategy for small image patches, so as to effectively detect local image manipulations.

\bibliographystyle{IEEEtran}
\bibliography{PapersList}

\end{document}